\documentclass[11pt,a4paper]{article}
\usepackage{authblk}
\usepackage{amsmath,amssymb}
\usepackage[dvips]{graphicx}
\usepackage{array}
\usepackage{comment}
\usepackage{authblk}
\usepackage{epstopdf}

\usepackage{subfig}
\usepackage{color}

\usepackage{color}
\usepackage{psfrag}
\usepackage{soul}
\usepackage{rotating}
\usepackage{multirow}
\usepackage{enumerate}
\usepackage{caption}
\usepackage[normalem]{ulem}
\usepackage[super]{natbib}

\title{Monitoring the Multivariate Coefficient of Variation using Run Rules Type Control Charts}
 \author[1,4]{P. H. Tran*} 
  \author[2]{A. C. Rakitzis} 
  \author[1]{H. D. Nguyen}
   \author[1]{Q. T. Nguyen}
    \author[3]{K. P. Tran}
     \author[4]{C. Heuchenne}
\affil[1]{Institute of Artificial Intelligence and Data Science, Dong A University, Da Nang, Vietnam}
   \affil[2]{Department of Statistics and Actuarial - Financial Mathematics, Karlovasi, Samos, Greece } 
  \affil[3]{Ecole Nationale Sup\'erieure des Arts et Industries Textiles, GEMTEX Laboratory, BP 30329 59056 Roubaix Cedex 1, France}
\affil[4]{HEC - Management School, University of Li\`ege, Li\`ege 4000, Belgium, {*}Corresponding author. Email:
 tran@doct.uliege.be}

\begin{document}

\maketitle

\begin{abstract}
In practice, there are processes where the in-control mean and standard deviation of a quality characteristic is not stable. In such cases, the coefficient of variation (CV) is a more appropriate measure for assessing process stability. In this paper, we consider the statistical design of Run Rules based control charts for monitoring the CV of multivariate data. A Markov chain approach is used to evaluate the statistical performance of the proposed charts. The computational results show that the Run Rules based charts outperform the standard Shewhart control chart significantly. Moreover, by choosing an appropriate scheme, the Run Rules based charts perform better than the Rum Sum control chart for monitoring the multivariate CV. An example in a spring manufacturing process is given to illustrate the implementation of the proposed charts. 
\end{abstract}




\section{Introduction}

\label{sec:introduction}
Representing the ratio of the standard deviation to the mean, the coefficient of variation (CV) is a useful measure of relative dispersion of a random variable. It has the meaning that the higher the CV, the greater the level of dispersion around the mean. The CV is widely used in a large number of areas such as clinical chemistry, materials engineering, agricultural experiments and medicine, see, for example \citet{Castagliola_EWMA_CV_2011}. In laboratory medicine for comparing the reproducibility of assay techniques, a lower CV leads  to a better analytical precision. In many manufacturing processes, keeping the CV in-control means ensuring the product quality. Monitoring the CV is then an important task red that had attracted the interest of many authors and the literature of monitoring the CV is abundant. \citet{Kang_CV_2007} are the first to use the Shewhart chart to monitor and detect changes in the CV. Their work is developed with a number of advantage-type control charts or adaptive strategies such as the synthetic control chart (\citet{Calzada2013}), EWMA control chart (\citet{Castagliola_EWMA_CV_2011}), CUSUM control chart (\citet{Tranhanh2016_CUSUM_CV}), Run Rules based control chart (\citet{Castagliola_Run-Rules_CV_2013}), variable sampling interval (VSI) control chart (\citet{CastagliolaVSI_CV_2013}), side-sensitive group runs (SSGR) control chart (\citet{You2015_CV_sensitive}), and Run Sum control chart (\citet{Teoh2017Runsum_CV}). The EWMA chart designed by \citet{Castagliola_EWMA_CV_2011} was further improved by \citet{Zhang2014} (based on a modified EWMA charting statistic), \citet{Yeong_VSI_EWMA_CV_2017} (which integrates the VSI feature into the EWMA chart) and Zhang et al. \cite{zhang2018control} (by applying the resetting technique in  \citet{shu2008new}).\\

The aforementioned studies focus on the univariate CV. In fact, there are various situations where the multivariate coefficient of variation (MCV) is the main concern. For example, in biometry and genetics, it is quite often to measure multiple characteristics on individuals from several populations and the problem is to assess the relative variability of each population. The single calculation of the univariate CV of each characteristic is obviously insufficient because it does not consider the correlation between these features, see \citet{albert2010novel}. However, a literature search reveals that not much attention has been paid to the CV for multivariate data despite its potential importance. Very recently, \citet{Yeong_MCV_2016} have suggested a Shewhart control chart for monitoring the MCV (denoted as Shewhart$-$MCV chart in this paper). It is well-known that Shewhart-type charts are only efficient in detecting large and sudden process shifts and it is not the case for moderate or small shifts. To enhance the performance of the Shewhart chart monitoring MCV in detecting small shifts, \citet{Lim2017runsum-MCV} suggested to use a Rum Sum control chart. In this paper, we propose to apply supplementary Run Rules with the Shewhart MCV control chart.\\

In the literature, several Run Rules charts have been studied  and suggested by a number of authors. 
\citet{champ1987exact} were the first to obtain the exact formula evaluating the run length distribution and then calculated its average value, i.e. the average run length (ARL) which is the expected number of points plotted on chart until an out-of-control signal is given. They have also showed a disadvantage of the use of supplementary runs rules in the Shewhart chart as it reduces the in-control $ARL$ (denoted as $ARL_0$). In order to overcome this downside, alternative methods for the statistical design of control charts with Run Rules have been proposed: \citet{Klein2000} considered the 2-of-2 and 2-of-3 schemes while \citet{Khoo2004} considered the 2-of-4, 3-of-3 and 3-of-4 schemes. A modified version of $r$-out-of-$m$ control chart was studied in \citet{Antzoulakos2008}. An overview of control charts with supplementary runs rules until 2006 is presented in Koutras et al. \cite{Koutras_2007}. Recently, the Run Rules control chart are applied to monitor the coefficient of variation, the ratio of two normal variables as well as other non-normal processes; see, for example \citet{Acosta_2009_kofkRR,Amdouni_2016_one-sided-RR,Castagliola_Run-Rules_CV_2013,Faraz_2014_T2RR}, Tran et al. \cite{Tran2016_Runrules_RZ}. Recently, \citet{chew2019efficiency} studied upper and lower one-sided run rules control charts, based on the 2-of-3 and the 3-of-4 runs rules, for the MCV. However, in this work we consider additionally the 4-of-5 runs rule, a rule that it is well-know for its efficiency (see, for example, \citet{Tran2016_4-out-of-5-Runrules_RZ}). Therefore, in this study we provide additional numerical comparisons between the most frequently used one-sided Shewhart-MCV charts with or without run rules. Numerical simulations shows that our proposed charts are efficient in detecting the process shifts. Moreover, the implementation of Run Rules control charts is also less complex compared to the Run Sum chart for MCV.\\

The paper is organized as follows. In Section \ref{sec:distribution}, we provide a brief review of distribution of the sample MCV as discovered by \citet{Yeong_MCV_2016}. In Section \ref{sec:implementationnonME}, we present the design and implementation of the Run Rules control charts for monitoring the MCV. Section \ref{sec:numerical} is devoted to assessing the performance of the proposed charts. An  example is illustrated in Section \ref{sec:illustrative} and some concluding remarks are given in Section \ref{sec:conclusions}.
\section{A brief review on the distribution of the sample multivariate coefficient of variation}
\label{sec:distribution}

We present in this Section a brief review of the distribution of the sample multivariate coefficient of variation (abbr. MCV). From the literature, there  are different point of views about the definitions of the MCV. A formal definition for the multivariate coefficient of variation was firstly proposed by \citet{Reyment1960}. \citet{van1974multivariate} has pointed out a few shortcomings of this definition and suggested a more appropriate definition. Another definition of the MCV was given by \citet{nikulin2011unbiased} based on the Mahalanobis distance. Recently, \citet{albert2010novel} proposed a novel approach of defining the MCV to overcome the limitations of previous definitions. We use in this paper the definition of the MCV suggested by \citet{nikulin2011unbiased} which is considered as a natural generalization for the CV. This definition was also adopted by \citet{Yeong_MCV_2016} to monitor the MCV. Let $\mathbf{X}$ denote a random vector from a $p$-variate normal distribution with mean vector $\boldsymbol{\mu}$ and covariance matrix $\boldsymbol{\Sigma}$. The MCV is defined as \begin{equation}
\label{equ:gamma}
  \gamma=(\boldsymbol{\mu}^T\boldsymbol{\Sigma}^{-1}\boldsymbol{\mu})^{-\frac{1}{2}}.
\end{equation} 

Suppose that a random sample of size $n$, say  $\mathbf{X}_1, \mathbf{X}_2,\ldots, \mathbf{X}_n$, is taken from this distribution, i.e., $\mathbf{X}_i=(x_{i,1},x_{i,2},\ldots,x_{i,p})\sim N(\boldsymbol{\mu},\boldsymbol{\Sigma})$, $i=1,\ldots,n$. Let $\bar{\mathbf{X}}$ and $\mathbf{S}$ be the sample mean and the sample covariance matrix of $\mathbf{X}_1, \mathbf{X}_2,\ldots, \mathbf{X}_n$,
i.e.,
\[
\bar{\mathbf{X}}=\frac{1}{n}\sum_{i=1}^n\mathbf{X}_i,
\]
\noindent
and
\[
\mathbf{S}=\frac{1}{n-1}\sum_{i=1}^n(\mathbf{X}_i-\bar{\mathbf{X}})(\mathbf{X}_i-\bar{\mathbf{X}})^T.
\]
The sample multivariate coefficient of variation $\hat{\gamma}$
is then defined as 
\begin{equation}
\hat{\gamma}_i=(\bar{\mathbf{X}}^T\mathbf{S}^{-1}\bar{\mathbf{X}})^{-\frac{1}{2}}.
\end{equation}

The $c.d.f$ (cumulative distribution function) and the inverse $c.d.f$ of $\hat{\gamma}$ are given in \citet{Yeong_MCV_2016} as 
\begin{equation}
  \label{equ:CDFMCV}
  F_{\hat{\gamma}}(x|n,p,\delta)=1-F_F\left(\frac{n(n-p)}{(n-1)px^2}|p,n-p,\delta\right),
\end{equation}
\noindent
and 
\begin{equation}
  \label{equ:IDFMCV}
  F^{-1}_{\hat{\gamma}}(\alpha|n,p,\delta)=\sqrt{\frac{n(n-p)}{(n-1)p}\left(\frac{1}{F^{-1}_F(1-\alpha|p,n-p,\delta)}\right)},
\end{equation}
\noindent
where $F_F(.|p,n-p,\delta)$ and $F^{-1}_F(.|p,n-p,\delta)$  are the non-central $F$ distribution and the inverse of the non-central $F$ distribution with $p$ and $n-p$ degrees of freedom, respectively, and the non-centrality parameter is $\delta=n\boldsymbol{\mu}^T\boldsymbol{\Sigma}^{-1}\boldsymbol{\mu}=\frac{n}{\gamma^2}$.

\section{Implementation of the RR$_{r,s}$-MCV control charts}
\label{sec:implementationnonME}
Similar to the one-sided Run Rules control charts presented in Tran et al.  \cite{Tran2016_Runrules_RZ} and \citet{Castagliola_Run-Rules_CV_2013}, we suggest the definition of two one-sided Run Rules  control charts for monitoring the MCV as follows:
\begin{itemize}
\item A lower-sided $r$-out-of-$s$ Run Rules control chart (denoted as
  RR$^-_{r,s}-$ MCV) to detect a decrease in $\hat{\gamma}_i$ with a
  lower control limit $LCL^-$ and a corresponding upper control
  limit $UCL^-=+\infty$. 
\item An upper-sided $r$-out-of-$s$ Run Rules control chart (denoted as
  RR$^+_{r,s}-$ MCV) to detect an increase in $\hat{\gamma}_i$ with an upper control limit $UCL^+$ and a corresponding lower control
  limit $LCL^+=0$.  
\end{itemize}

Given the value of the control limits for each chart, an out-of-control signal is given at time $i$ if $r$-out-of-$s$ consecutive $\hat{\gamma}_i$  values are plotted outside the control interval, i.e. $\hat{\gamma}_i<LCL^-$ in the lower-sided chart and $\hat{\gamma}_i>UCL^+$ in the upper-sided chart. The control chart designed above is called $pure$ Run Rules type chart. Compared to the $composite$ Run Rules type charts which require both control and warning limits, these $pure$ type charts are more simple to implement and interpret, see \citet{Klein2000}. In this study, we only consider the 2-out-of-3, 3-out-of-4 and 4-out-of-5 Run Rules charts. More complex Run Rules schemes with larger values of $(r,s)$ are possible to design in a similar manner. However, their efficiency should be taken into consideration in terms of the increased complexity of implementation.

\vspace{0.25cm}
The performance of the proposed one-sided RR$_{r,s}-$MCV control charts is measured by the out-of-control $ARL$,  denoted as $ARL_1$.  We utilize a Markov chain method, similar to the one initially proposed by  \citet{Brook1972}, to calculate the $ARL_1$ value. Further details on this method can be found in Fu et al. \cite{Fu_2003_Markov_approach}, \citet{Castagliola_Run-Rules_CV_2013} and \citet{Li2014}. 
Let us now suppose that the occurrence of an unexpected condition shifts the in-control MCV value $\gamma_0$  to the out-of-control value $\gamma_1=\tau\times \gamma_0$, where $\tau>0$ is the shift size. Values of $\tau\in (0,1)$ correspond to a decrease of the $\gamma_0$, while values of $\tau>1$ correspond to an increase of the in-control MCV. It is worth mentioning that a decrease (resp. increase) in $\gamma_0$ is related to process improvement (resp. deterioration). The probability $p$ of the event that  a sample falls into an in-control interval is equal to:
\begin{itemize}
\item for the  RR$_{r,s}^--$MCV chart:
  \begin{equation}
  p=P(\hat{\gamma}_i\geq LCL^-)= 1- F_{\hat{\gamma}}(LCL^-|n,p,\delta_1),
  \end{equation}
\item for the  RR$_{r,s}^+-$MCV chart:
  \begin{equation}
  p=P(\hat{\gamma}_i\leq UCL^+)= F_{\hat{\gamma}}(UCL^+|n,p,\delta_1),
  \end{equation}
\end{itemize}
where $\delta_1=\frac{n}{{(\tau \gamma_0)}^2}$.\\

The Transition Probability Matrix (TPM) matrix  $\mathbf{P}$ of the embedded Markov chain for the two RR$_{2,3}-$MCV control charts is
\begin{equation}
\label{equ:P}
\mathbf{P}=\left(
  \begin{array}{cc}
  \mathbf{Q} & \mathbf{r} \\
  & \\
  \mathbf{0}^T & 1
  \end{array}
\right)
=\left(
  \begin{array}{ccc|c}
  0 & 0 & p & 1-p \\
  p & 0 & 0 & 1-p \\
  0 & 1-p & p & 0 \\
  \hline
  0 & 0 & 0 & 1
  \end{array}
\right),
\end{equation}
where $\mathbf{Q}$ is the $(3,3)$ matrix of
transient probabilities, $\mathbf{r}$ is the $(3,1)$ vector satisfied
$\mathbf{r}=\mathbf{1}-\mathbf{Q}\mathbf{1}$ with $\mathbf{1}=(1,1,1)^T$ and $\mathbf{0}=(0,0,0)^T$. The corresponding $(3,1)$ vector $\mathbf{q}$ of initial probabilities associated with the transient states is equal to $\mathbf{q}=(0,0,1)^T$ , i.e. the third state is the initial state.

\vspace{0.25cm}
Extended to Run Rules charts with larger $(r,s)$ values, the matrix
$\mathbf{Q}_{(7\times 7)}$ of transient probabilities for the two RR$_{3,4}-$MCV control charts is given by
\begin{equation}
\mathbf{Q}=\left(
\begin{array}{ccccccc}
0 & 0 & p & 0 & 0 & 0 & 0 \\
0 & 0 & 0 & 0 & p & 0 & 0 \\
0 & 0 & 0 & 0 & 0 & 1-p & p \\
p & 0 & 0 & 0 & 0 & 0 & 0 \\
0 & 1-p & p & 0 & 0 & 0 & 0 \\
0 & 0 & 0 & 1-p & p & 0 & 0 \\
0 & 0 & 0 & 0 & 0 & 1-p & p \\
\end{array}
\right).
\end{equation}
In this case, the seventh state in the vector $\mathbf{q}=(0,0,0,0,0,0,1)^T$  is the initial state. The (15,15) matrix $\mathbf{Q}$ of  transient probabilities for the two RR$_{4,5}$-MCV control charts is not presented here due to space constraints, but it can be seen in, for example, \citet{Tran2016_Runrules_RZ}. Once matrix $\mathbf{Q}$ and vector $\mathbf{q}$ have been determined, the $ARL$ and $SDRL$ (standard deviation of run length) are given by 

%

\begin{eqnarray}
\label{ARL}
ARL & = & \nu_1, \\
SDRL & = & \sqrt{\mu_2},
\end{eqnarray}

with
\begin{eqnarray}
\nu_1 & = & \mathbf{q}^T(\mathbf{I}-\mathbf{Q})^{-1}\mathbf{1},\\
\nu_2 & = & 2\mathbf{q}^T(\mathbf{I}-\mathbf{Q})^{-2}\mathbf{Q}\mathbf{1},\\
\mu_2 & = & \nu_2-\nu_1^2+\nu_1.
\end{eqnarray}

 A control chart is considered to be better than its competitors if it gives smaller value of the $ARL_1$ while the $ARL_0$ is  the same. Therefore, the control limit of the RR$_{r,s}-$MCV control charts should be found out as a solution of the following equations:
\begin{itemize}
\item for the  RR$_{r,s}^--$MCV chart:
  \begin{equation} \label{equ:control limit 1}
  ARL(LCL^-,n,p,\gamma_0,\tau=1)=ARL_0,
  \end{equation}
\item for the  RR$_{r,s}^+-$MCV chart:
  \begin{equation} \label{equ:control limit 2}
 ARL(UCL^+,n,p,\gamma_0,\tau=1)=ARL_0,
  \end{equation}
\end{itemize}
where $ARL_0$ is the predetermined in-control $ARL$ value.
\section{The Performance of the RR$_{r,s}$-MCV control charts}
\label{sec:numerical}
In this Section, we investigate the performance of the RR $_{r,s}-$ MCV control charts.  The desired in-control $ARL$ value, say $ARL_0$ is set at 370.4, for all the considered IC cases. The control limit $LCL^-$ of lower-sided chart and $UCL ^+$ of upper-sided chart, which are the solution of equations  (\ref{equ:control limit 1}) and (\ref{equ:control limit 2}),  are shown in Table~\ref{tab:tabULCL} for different combinations of $n\in \{5,10,15\}$, $p\in\{2, 3, 4\}$ and $\gamma_0\in \{0.1$, $0.2$, $0.3$, $0.4$, $0.5\}$.

\begin{table}[thb]
  \hspace*{-18mm}
  \scalebox{0.62}{
  \begin{tabular}{cccccccccccc}
   \hline
  & \multicolumn{3}{c}{RR$_{2,3}$-MCV chart} & &\multicolumn{3}{c}{RR$_{3,4}$-MCV chart} && \multicolumn{3}{c}{RR$_{4,5}$-MCV chart}\\
  \cline{2-4} \cline{6-8} \cline{10-12}
$\gamma_0$&  $n=5$ & $n=10$ & $n=15$  &&  $n=5$ & $n=10$ & $n=15$ &&  $n=5$ & $n=10$ & $n=15$ \\
     \cline{2-4} \cline{6-8} \cline{10-12}
        &  \multicolumn{11}{c}{$p=2$} \\
0.10 & (0.027, 0.146) & (0.053, 0.135) & (0.063, 0.129) & & (0.039, 0.124) & (0.063, 0.121) & (0.071, 0.119) & & (0.048, 0.111) & (0.070, 0.113) & (0.077, 0.112) \\
0.20 & (0.053, 0.296) & (0.104, 0.274) & (0.125, 0.261) & & (0.077, 0.251) & (0.125, 0.245) & (0.142, 0.239) & & (0.095, 0.223) & (0.139, 0.227) & (0.154, 0.224) \\
0.30 & (0.079, 0.457) & (0.155, 0.419) & (0.185, 0.399) & & (0.114, 0.382) & (0.185, 0.372) & (0.211, 0.362) & & (0.142, 0.337) & (0.206, 0.343) & (0.229, 0.339) \\
0.40 & (0.104, 0.633) & (0.203, 0.574) & (0.242, 0.544) & & (0.150, 0.520) & (0.243, 0.504) & (0.277, 0.490) & & (0.186, 0.455) & (0.272, 0.462) & (0.302, 0.456) \\
0.50 & (0.127, 0.831) & (0.248, 0.744) & (0.297, 0.700) & & (0.184, 0.667) & (0.299, 0.643) & (0.342, 0.623) & & (0.229, 0.576) & (0.335, 0.584) & (0.373, 0.577) \\

    \cline{2-4} \cline{6-8} \cline{10-12}
    &  \multicolumn{11}{c}{$p=3$} \\
    0.10 & (0.014, 0.128) & (0.047, 0.129) & (0.059, 0.125) & & (0.024, 0.106) & (0.057, 0.115) & (0.067, 0.115) & & (0.031, 0.093) & (0.063, 0.106) & (0.073, 0.108) \\
0.20 & (0.028, 0.259) & (0.092, 0.260) & (0.117, 0.253) & & (0.047, 0.213) & (0.112, 0.231) & (0.134, 0.231) & & (0.062, 0.185) & (0.126, 0.213) & (0.146, 0.216) \\
0.30 & (0.041, 0.395) & (0.137, 0.397) & (0.173, 0.385) & & (0.069, 0.322) & (0.166, 0.350) & (0.199, 0.349) & & (0.092, 0.279) & (0.187, 0.321) & (0.217, 0.326) \\
0.40 & (0.054, 0.539) & (0.179, 0.541) & (0.227, 0.524) & & (0.090, 0.434) & (0.218, 0.472) & (0.262, 0.470) & & (0.121, 0.372) & (0.246, 0.431) & (0.286, 0.438) \\
0.50 & (0.065, 0.692) & (0.219, 0.695) & (0.278, 0.670) & & (0.110, 0.547) & (0.267, 0.599) & (0.322, 0.596) & & (0.148, 0.465) & (0.303, 0.543) & (0.352, 0.551) \\
    \cline{2-4} \cline{6-8} \cline{10-12}
       &  \multicolumn{11}{c}{$p=4$} \\
    0.10 & (0.002, 0.104) & (0.040, 0.122) & (0.055, 0.121) & & (0.007, 0.081) & (0.050, 0.108) & (0.063, 0.111) & & (0.011, 0.067) & (0.057, 0.099) & (0.069, 0.104) \\
0.20 & (0.005, 0.208) & (0.080, 0.246) & (0.109, 0.245) & & (0.013, 0.162) & (0.099, 0.217) & (0.126, 0.222) & & (0.023, 0.133) & (0.113, 0.199) & (0.138, 0.208) \\
0.30 & (0.007, 0.314) & (0.118, 0.373) & (0.161, 0.371) & & (0.019, 0.242) & (0.146, 0.327) & (0.187, 0.335) & & (0.033, 0.199) & (0.167, 0.299) & (0.204, 0.313) \\
0.40 & (0.009, 0.421) & (0.154, 0.506) & (0.211, 0.503) & & (0.025, 0.321) & (0.192, 0.440) & (0.245, 0.451) & & (0.044, 0.262) & (0.219, 0.399) & (0.269, 0.418) \\
0.50 & (0.011, 0.529) & (0.188, 0.645) & (0.259, 0.641) & & (0.031, 0.399) & (0.234, 0.553) & (0.301, 0.569) & & (0.053, 0.324) & (0.268, 0.500) & (0.331, 0.525) \\
\hline
    \end{tabular}}
   \caption{ The control limits of RR$_{r,s}$-MCV control charts,  $LCL^-$ on left side  and $UCL^+$ on right side, for different values of $n,p$ and $\gamma_0$.}
  \label{tab:tabULCL}
  \end{table}
  
It can be seen from Table \ref{tab:tabULCL} that  given $n$ and $p$ the values of $LCL^-$ and $UCL^+$ depend (in general) on $\gamma_0$ .  In particular, small values of $\gamma_0$ lead to  small values of $LCL^-$ and $UCL^+$, as well. For example, in RR$_{2,3}-$MCV chart with $n=5$ and $p=2$, we have $LCL^-=0.027$  and $UCL^+=0.146$ when $\gamma_0=0.10$, while $LCL^-=0.127$  and $UCL^+=0.831$ when  $\gamma_0=0.50$. Also, given $n$ and $\gamma_0$, the values of $LCL^-$ and $UCL^+$ depend on $p$.  The larger the value of $p$, the smaller the value of $LCL^-$ and $UCL^+$. For example, in RR$_{3,4}-$MCV chart  with $n=10$ and $\gamma_0=0.2$, we have $LCL^-=0.125$ and  $UCL^+=0.245$ when $p=2$, while $LCL^-=0.099$ and  $UCL^+=0.217$ when $p=4$.
   
\vspace{0.25cm}
Using the control limits in Table \ref{tab:tabULCL}, the values of $ARL_1$ and $SDRL_1$ (out-of-control) of the RR$_{r,s}-$MCV control charts are provided in Tables~\ref{tab:tabARL1}-\ref{tab:tabARL3}. We set different combinations of $\tau\in \{0.50$, $0.75, 0.9,1.10,1.25,1.5\}$, $p\in \{2, 3, 4\}$, $n\in\{5,10,15\}$ and $\gamma_0\in\{0.1, 0.2, 0.3, 0.4, 0.5\}$. Some simple conclusions can be drawn from these tables as follows: 
\begin{table}[thb]
  \hspace*{-15mm}
  \scalebox{0.64}{
    \begin{tabular}{cccccccccccc}
     \hline
  & \multicolumn{3}{c}{RR$_{2,3}$-MCV chart} & &\multicolumn{3}{c}{RR$_{3,4}$-MCV chart} && \multicolumn{3}{c}{RR$_{4,5}$-MCV chart}\\
      \cline{2-4} \cline{6-8} \cline{10-12}
$\tau$&  $n=5$ & $n=10$ & $n=15$  &&  $n=5$ & $n=10$ & $n=15$ &&  $n=5$ & $n=10$ & $n=15$ \\
     \cline{2-4} \cline{6-8} \cline{10-12}
        &  \multicolumn{11}{c}{$\gamma_0=0.1$} \\

0.50 & (14.2, 12.6) & (2.8, 1.3) & (2.1, 0.4) & & (8.5, 6.1) & (3.3, 0.7) & (3.0, 0.2) & & (7.1, 3.9) & (4.1, 0.4) & (4.0, 0.1) \\
0.75 & (84.0, 82.2) & (21.6, 20.0) & (10.4, 8.9) & & (55.6, 52.9) & (14.8, 12.4) & (8.0, 5.7) & & (42.2, 38.9) & (12.4, 9.3) & (7.5, 4.4) \\
0.90 & (211.8, 210.0) & (116.2, 114.4) & (77.7, 75.9) & & (177.0, 174.2) & (90.3, 87.6) & (58.9, 56.3) & & (154.8, 151.1) & (76.6, 73.1) & (49.8, 46.4) \\
1.10 & (109.6, 107.7) & (67.0, 65.2) & (48.9, 47.1) & & (109.3, 106.5) & (63.4, 60.8) & (45.2, 42.6) & & (111.1, 107.5) & (62.6, 59.1) & (44.1, 40.7) \\
1.25 & (32.5, 30.8) & (14.7, 13.0) & (9.4, 7.9) & & (33.5, 30.9) & (14.7, 12.3) & (9.6, 7.2) & & (35.3, 32.0) & (15.3, 12.2) & (10.2, 7.1) \\
1.50 & (10.5, 8.9) & (4.6, 3.1) & (3.3, 1.7) & & (11.7, 9.3) & (5.4, 3.0) & (4.0, 1.6) & & (13.1, 10.0) & (6.3, 3.1) & (4.9, 1.6) \\
     \cline{2-4} \cline{6-8} \cline{10-12}
        &  \multicolumn{11}{c}{$\gamma_0=0.2$} \\
0.50 & (14.6, 13.0) & (2.9, 1.3) & (2.1, 0.4) & & (8.7, 6.3) & (3.3, 0.7) & (3.0, 0.2) & & (7.2, 4.0) & (4.1, 0.4) & (4.0, 0.1) \\
0.75 & (85.6, 83.8) & (22.5, 20.9) & (10.9, 9.3) & & (56.8, 54.2) & (15.4, 12.9) & (8.3, 6.0) & & (43.3, 39.9) & (12.9, 9.8) & (7.8, 4.7) \\
0.90 & (213.7, 211.8) & (118.9, 117.1) & (80.2, 78.4) & & (179.0, 176.2) & (92.8, 90.0) & (61.0, 58.4) & & (156.9, 153.2) & (78.8, 75.3) & (51.7, 48.3) \\
1.10 & (111.9, 110.1) & (69.5, 67.8) & (51.1, 49.4) & & (111.6, 108.8) & (65.8, 63.1) & (47.3, 44.7) & & (113.3, 109.7) & (64.9, 61.4) & (46.1, 42.7) \\
1.25 & (33.8, 32.1) & (15.6, 13.9) & (10.1, 8.5) & & (34.8, 32.2) & (15.5, 13.1) & (10.1, 7.8) & & (36.6, 33.3) & (16.1, 13.0) & (10.7, 7.6) \\
1.50 & (11.1, 9.5) & (4.9, 3.4) & (3.4, 1.9) & & (12.3, 9.9) & (5.7, 3.3) & (4.2, 1.8) & & (13.7, 10.5) & (6.5, 3.4) & (5.1, 1.8) \\
     \cline{2-4} \cline{6-8} \cline{10-12}
        &  \multicolumn{11}{c}{$\gamma_0=0.3$} \\
0.50 & (15.2, 13.6) & (3.0, 1.5) & (2.2, 0.5) & & (9.0, 6.6) & (3.4, 0.8) & (3.0, 0.2) & & (7.4, 4.3) & (4.1, 0.5) & (4.0, 0.1) \\
0.75 & (88.2, 86.4) & (24.0, 22.3) & (11.7, 10.2) & & (58.9, 56.3) & (16.3, 13.9) & (8.9, 6.5) & & (45.0, 41.6) & (13.7, 10.5) & (8.2, 5.1) \\
0.90 & (216.6, 214.8) & (123.1, 121.3) & (84.1, 82.3) & & (182.3, 179.5) & (96.6, 93.9) & (64.5, 61.8) & & (160.2, 156.5) & (82.4, 78.9) & (54.7, 51.3) \\
1.10 & (116.0, 114.2) & (73.8, 72.0) & (54.9, 53.2) & & (115.5, 112.7) & (69.8, 67.1) & (50.7, 48.1) & & (117.2, 113.5) & (68.7, 65.2) & (49.4, 46.0) \\
1.25 & (36.2, 34.4) & (17.1, 15.5) & (11.1, 9.6) & & (37.0, 34.4) & (16.9, 14.5) & (11.1, 8.7) & & (38.9, 35.5) & (17.5, 14.3) & (11.6, 8.5) \\
1.50 & (12.1, 10.5) & (5.4, 3.9) & (3.7, 2.2) & & (13.3, 10.9) & (6.1, 3.8) & (4.5, 2.1) & & (14.7, 11.6) & (7.0, 3.8) & (5.3, 2.1) \\
      \cline{2-4} \cline{6-8} \cline{10-12}
        &  \multicolumn{11}{c}{$\gamma_0=0.4$} \\
0.50 & (16.0, 14.4) & (3.1, 1.6) & (2.2, 0.5) & & (9.4, 7.1) & (3.4, 0.9) & (3.1, 0.2) & & (7.7, 4.6) & (4.2, 0.5) & (4.0, 0.1) \\
0.75 & (91.7, 89.9) & (26.0, 24.3) & (12.9, 11.3) & & (61.8, 59.2) & (17.7, 15.3) & (9.6, 7.3) & & (47.4, 44.0) & (14.7, 11.6) & (8.8, 5.7) \\
0.90 & (220.5, 218.7) & (128.5, 126.7) & (89.3, 87.5) & & (186.6, 183.8) & (101.8, 99.0) & (69.0, 66.3) & & (164.6, 160.9) & (87.2, 83.7) & (58.8, 55.3) \\
1.10 & (121.9, 120.1) & (79.6, 77.8) & (60.1, 58.4) & & (121.0, 118.2) & (75.2, 72.5) & (55.4, 52.8) & & (122.5, 118.9) & (73.9, 70.4) & (53.9, 50.4) \\
1.25 & (39.7, 38.0) & (19.4, 17.8) & (12.7, 11.2) & & (40.3, 37.7) & (18.9, 16.5) & (12.5, 10.1) & & (42.1, 38.7) & (19.4, 16.2) & (12.9, 9.8) \\
1.50 & (13.7, 12.1) & (6.2, 4.7) & (4.2, 2.7) & & (14.8, 12.4) & (6.8, 4.5) & (4.9, 2.5) & & (16.2, 13.1) & (7.6, 4.5) & (5.7, 2.5) \\

     \cline{2-4} \cline{6-8} \cline{10-12}
        &  \multicolumn{11}{c}{$\gamma_0=0.5$} \\
        
        0.50 & (17.1, 15.5) & (3.3, 1.8) & (2.3, 0.6) & & (10.0, 7.7) & (3.5, 1.0) & (3.1, 0.3) & & (8.1, 5.0) & (4.2, 0.6) & (4.0, 0.2) \\
0.75 & (96.1, 94.3) & (28.6, 26.9) & (14.4, 12.8) & & (65.5, 62.8) & (19.5, 17.0) & (10.6, 8.3) & & (50.5, 47.1) & (16.1, 12.9) & (9.6, 6.5) \\
0.90 & (225.1, 223.3) & (134.9, 133.1) & (95.5, 93.7) & & (191.7, 188.9) & (107.8, 105.1) & (74.5, 71.8) & & (170.0, 166.3) & (93.0, 89.4) & (63.8, 60.3) \\
1.10 & (129.7, 127.9) & (87.0, 85.2) & (66.7, 64.9) & & (128.1, 125.3) & (81.9, 79.2) & (61.3, 58.7) & & (129.2, 125.6) & (80.3, 76.8) & (59.4, 55.9) \\
1.25 & (44.7, 43.0) & (22.5, 20.9) & (14.9, 13.3) & & (44.8, 42.2) & (21.6, 19.2) & (14.3, 11.9) & & (46.4, 43.0) & (22.0, 18.7) & (14.7, 11.5) \\
1.50 & (16.0, 14.4) & (7.3, 5.7) & (4.8, 3.3) & & (16.9, 14.5) & (7.8, 5.4) & (5.4, 3.1) & & (18.3, 15.1) & (8.5, 5.4) & (6.2, 3.0) \\
\hline
    \end{tabular}}
   \caption{The values of $ARL_1$ and $SDRL_1$ using the RR$_{r,s}-$MCV control charts for \textbf{$p=2$} and different values of  $\tau,n$ and $\gamma_0$.}
  \label{tab:tabARL1}
  \end{table}

\begin{table}[thb]
  \hspace*{-15mm}
  \scalebox{0.64}{
    \begin{tabular}{cccccccccccc}
     \hline
  & \multicolumn{3}{c}{RR$_{2,3}$-MCV chart} & &\multicolumn{3}{c}{RR$_{3,4}$-MCV chart} && \multicolumn{3}{c}{RR$_{4,5}$-MCV chart}\\
      \cline{2-4} \cline{6-8} \cline{10-12}
$\tau$&  $n=5$ & $n=10$ & $n=15$  &&  $n=5$ & $n=10$ & $n=15$ &&  $n=5$ & $n=10$ & $n=15$ \\
     \cline{2-4} \cline{6-8} \cline{10-12}
        &  \multicolumn{11}{c}{$\gamma_0=0.1$} \\
        0.50 & (32.7, 31.0) & (3.3, 1.8) & (2.2, 0.5) & & (17.4, 15.0) & (3.5, 0.9) & (3.0, 0.2) & & (12.4, 9.2) & (4.2, 0.6) & (4.0, 0.1) \\
0.75 & (129.4, 127.6) & (26.5, 24.8) & (11.7, 10.2) & & (90.5, 87.8) & (17.7, 15.3) & (8.8, 6.5) & & (69.5, 66.0) & (14.6, 11.4) & (8.1, 5.0) \\
0.90 & (250.7, 248.9) & (128.2, 126.4) & (83.4, 81.6) & & (217.2, 214.4) & (100.5, 97.8) & (63.5, 60.8) & & (193.7, 190.0) & (85.5, 81.9) & (53.7, 50.2) \\
1.10 & (128.4, 126.6) & (72.4, 70.7) & (51.7, 49.9) & & (130.8, 128.0) & (69.1, 66.4) & (48.0, 45.3) & & (134.6, 130.9) & (68.4, 64.9) & (46.9, 43.5) \\
1.25 & (43.1, 41.3) & (16.5, 14.9) & (10.1, 8.6) & & (45.6, 43.0) & (16.5, 14.1) & (10.3, 7.9) & & (48.8, 45.4) & (17.2, 14.0) & (10.9, 7.8) \\
1.50 & (14.8, 13.2) & (5.1, 3.6) & (3.4, 1.9) & & (16.7, 14.3) & (5.9, 3.6) & (4.2, 1.8) & & (18.8, 15.6) & (6.8, 3.7) & (5.1, 1.8) \\

     \cline{2-4} \cline{6-8} \cline{10-12}
        &  \multicolumn{11}{c}{$\gamma_0=0.2$} \\
0.50 & (33.4, 31.7) & (3.4, 1.8) & (2.2, 0.5) & & (17.8, 15.4) & (3.5, 1.0) & (3.0, 0.2) & & (12.6, 9.5) & (4.2, 0.6) & (4.0, 0.1) \\
0.75 & (131.2, 129.4) & (27.5, 25.8) & (12.3, 10.7) & & (92.1, 89.4) & (18.4, 15.9) & (9.2, 6.8) & & (70.9, 67.4) & (15.1, 11.9) & (8.4, 5.3) \\
0.90 & (252.3, 250.4) & (131.0, 129.1) & (86.0, 84.2) & & (219.0, 216.2) & (103.0, 100.3) & (65.7, 63.0) & & (195.7, 192.0) & (87.8, 84.3) & (55.6, 52.2) \\
1.10 & (130.9, 129.0) & (75.1, 73.3) & (54.0, 52.3) & & (133.2, 130.4) & (71.6, 68.9) & (50.1, 47.5) & & (137.0, 133.3) & (70.8, 67.3) & (49.0, 45.6) \\
1.25 & (44.7, 43.0) & (17.5, 15.9) & (10.8, 9.3) & & (47.2, 44.6) & (17.4, 15.0) & (10.9, 8.5) & & (50.5, 47.0) & (18.1, 14.9) & (11.5, 8.3) \\
1.50 & (15.6, 14.0) & (5.5, 4.0) & (3.6, 2.1) & & (17.5, 15.1) & (6.2, 3.9) & (4.4, 2.0) & & (19.7, 16.5) & (7.1, 4.0) & (5.3, 2.0) \\
     \cline{2-4} \cline{6-8} \cline{10-12}
        &  \multicolumn{11}{c}{$\gamma_0=0.3$} \\
0.50 & (34.5, 32.8) & (3.5, 2.0) & (2.2, 0.6) & & (18.5, 16.1) & (3.6, 1.1) & (3.1, 0.3) & & (13.1, 9.9) & (4.3, 0.7) & (4.0, 0.1) \\
0.75 & (134.2, 132.4) & (29.2, 27.5) & (13.2, 11.6) & & (94.8, 92.1) & (19.6, 17.1) & (9.8, 7.4) & & (73.3, 69.8) & (16.0, 12.8) & (8.9, 5.8) \\
0.90 & (254.8, 252.9) & (135.3, 133.4) & (90.1, 88.3) & & (222.0, 219.1) & (107.1, 104.3) & (69.3, 66.6) & & (198.9, 195.2) & (91.6, 88.1) & (58.8, 55.4) \\
1.10 & (135.1, 133.2) & (79.5, 77.7) & (57.9, 56.2) & & (137.2, 134.4) & (75.7, 73.0) & (53.7, 51.1) & & (141.0, 137.3) & (74.9, 71.3) & (52.4, 49.0) \\
1.25 & (47.6, 45.8) & (19.2, 17.6) & (12.0, 10.4) & & (50.1, 47.4) & (19.0, 16.6) & (11.9, 9.5) & & (53.3, 49.9) & (19.7, 16.4) & (12.5, 9.3) \\
1.50 & (17.0, 15.4) & (6.1, 4.6) & (4.0, 2.5) & & (19.0, 16.5) & (6.8, 4.4) & (4.7, 2.3) & & (21.2, 18.0) & (7.7, 4.5) & (5.5, 2.3) \\
     \cline{2-4} \cline{6-8} \cline{10-12}
        &  \multicolumn{11}{c}{$\gamma_0=0.4$} \\
0.50 & (36.2, 34.5) & (3.7, 2.2) & (2.3, 0.6) & & (19.5, 17.0) & (3.7, 1.2) & (3.1, 0.3) & & (13.7, 10.6) & (4.3, 0.7) & (4.0, 0.2) \\
0.75 & (138.2, 136.4) & (31.6, 29.9) & (14.5, 12.9) & & (98.5, 95.8) & (21.2, 18.7) & (10.6, 8.3) & & (76.6, 73.0) & (17.3, 14.1) & (9.5, 6.4) \\
0.90 & (258.1, 256.2) & (140.9, 139.0) & (95.5, 93.7) & & (225.9, 223.1) & (112.5, 109.7) & (74.0, 71.3) & & (203.1, 199.4) & (96.7, 93.1) & (63.1, 59.7) \\
1.10 & (141.2, 139.3) & (85.6, 83.8) & (63.4, 61.6) & & (143.0, 140.2) & (81.4, 78.7) & (58.6, 56.0) & & (146.5, 142.9) & (80.3, 76.8) & (57.1, 53.6) \\
1.25 & (51.9, 50.2) & (21.8, 20.2) & (13.7, 12.2) & & (54.2, 51.6) & (21.3, 18.8) & (13.4, 11.0) & & (57.5, 54.0) & (21.9, 18.6) & (13.9, 10.7) \\
1.50 & (19.3, 17.6) & (7.0, 5.4) & (4.5, 3.0) & & (21.2, 18.7) & (7.6, 5.3) & (5.1, 2.8) & & (23.5, 20.2) & (8.5, 5.3) & (6.0, 2.8) \\

     \cline{2-4} \cline{6-8} \cline{10-12}
        &  \multicolumn{11}{c}{$\gamma_0=0.5$} \\
0.50 & (38.3, 36.6) & (4.0, 2.5) & (2.4, 0.7) & & (20.8, 18.3) & (3.8, 1.3) & (3.1, 0.4) & & (14.6, 11.4) & (4.4, 0.9) & (4.0, 0.2) \\
0.75 & (143.2, 141.4) & (34.6, 32.9) & (16.2, 14.6) & & (103.2, 100.5) & (23.3, 20.8) & (11.8, 9.4) & & (80.8, 77.2) & (18.9, 15.7) & (10.4, 7.3) \\
0.90 & (262.1, 260.3) & (147.4, 145.6) & (101.9, 100.1) & & (230.7, 227.9) & (118.8, 116.1) & (79.8, 77.1) & & (208.4, 204.7) & (102.8, 99.2) & (68.4, 64.9) \\
1.10 & (149.3, 147.5) & (93.4, 91.6) & (70.2, 68.4) & & (150.5, 147.7) & (88.5, 85.8) & (64.8, 62.1) & & (153.7, 150.0) & (87.1, 83.6) & (62.9, 59.4) \\
1.25 & (58.0, 56.3) & (25.3, 23.6) & (16.1, 14.5) & & (59.9, 57.2) & (24.4, 21.9) & (15.5, 13.0) & & (63.0, 59.5) & (24.8, 21.5) & (15.8, 12.6) \\
1.50 & (22.6, 21.0) & (8.2, 6.7) & (5.2, 3.7) & & (24.4, 21.9) & (8.7, 6.4) & (5.8, 3.4) & & (26.7, 23.4) & (9.5, 6.4) & (6.6, 3.4) \\
\hline
    \end{tabular}}
   \caption{The values of $ARL_1$ and $SDRL_1$ using the RR$_{r,s}-$MCV control charts for \textbf{$p=3$} and different values of  $\tau,n$ and $\gamma_0$.}
  \label{tab:tabARL2}
  \end{table}

\begin{table}
 \hspace*{-15mm}
  \scalebox{0.64}{
    \begin{tabular}{cccccccccccc}
     \hline
  & \multicolumn{3}{c}{RR$_{2,3}$-MCV chart} & &\multicolumn{3}{c}{RR$_{3,4}$-MCV chart} && \multicolumn{3}{c}{RR$_{4,5}$-MCV chart}\\
      \cline{2-4} \cline{6-8} \cline{10-12}
$\tau$&  $n=5$ & $n=10$ & $n=15$  &&  $n=5$ & $n=10$ & $n=15$ &&  $n=5$ & $n=10$ & $n=15$ \\
     \cline{2-4} \cline{6-8} \cline{10-12}
        &  \multicolumn{11}{c}{$\gamma_0=0.1$} \\
        
0.50 & (101.5, 99.6) & (4.0, 2.5) & (2.3, 0.6) & & (61.8, 59.1) & (3.8, 1.3) & (3.1, 0.3) & & (41.7, 38.3) & (4.4, 0.8) & (4.0, 0.1) \\
0.75 & (215.1, 213.3) & (33.3, 31.6) & (13.4, 11.8) & & (172.2, 169.3) & (21.9, 19.4) & (9.8, 7.4) & & (141.9, 138.2) & (17.6, 14.4) & (8.8, 5.7) \\
0.90 & (303.4, 301.5) & (142.7, 140.9) & (90.0, 88.2) & & (278.9, 276.1) & (113.0, 110.3) & (68.7, 66.0) & & (258.8, 255.1) & (96.5, 92.9) & (58.1, 54.7) \\
1.10 & (161.2, 159.4) & (78.9, 77.1) & (54.8, 53.1) & & (169.6, 166.8) & (75.8, 73.1) & (51.1, 48.5) & & (178.0, 174.3) & (75.5, 71.9) & (50.0, 46.6) \\
1.25 & (65.9, 64.2) & (18.8, 17.2) & (11.0, 9.4) & & (73.5, 70.8) & (18.9, 16.4) & (11.1, 8.7) & & (81.1, 77.5) & (19.7, 16.5) & (11.7, 8.6) \\
1.50 & (25.8, 24.1) & (5.8, 4.3) & (3.6, 2.1) & & (30.7, 28.1) & (6.6, 4.3) & (4.4, 2.0) & & (35.6, 32.2) & (7.6, 4.4) & (5.3, 2.0) \\
    \cline{2-4} \cline{6-8} \cline{10-12}
        &  \multicolumn{11}{c}{$\gamma_0=0.2$} \\
0.50 & (102.7, 100.9) & (4.1, 2.6) & (2.3, 0.6) & & (62.8, 60.1) & (3.8, 1.4) & (3.1, 0.3) & & (42.5, 39.1) & (4.4, 0.9) & (4.0, 0.2) \\
0.75 & (216.7, 214.9) & (34.5, 32.8) & (14.0, 12.4) & & (173.9, 171.1) & (22.7, 20.2) & (10.2, 7.8) & & (143.7, 140.0) & (18.2, 15.0) & (9.2, 6.0) \\
0.90 & (304.3, 302.4) & (145.5, 143.7) & (92.6, 90.8) & & (280.2, 277.3) & (115.7, 112.9) & (71.1, 68.4) & & (260.3, 256.6) & (98.9, 95.4) & (60.2, 56.7) \\
1.10 & (163.7, 161.8) & (81.7, 79.9) & (57.3, 55.5) & & (172.0, 169.2) & (78.5, 75.8) & (53.4, 50.7) & & (180.4, 176.7) & (78.0, 74.5) & (52.2, 48.8) \\
1.25 & (68.0, 66.3) & (20.0, 18.3) & (11.7, 10.2) & & (75.6, 72.9) & (19.9, 17.5) & (11.7, 9.4) & & (83.3, 79.8) & (20.7, 17.5) & (12.3, 9.2) \\
1.50 & (27.1, 25.4) & (6.2, 4.7) & (3.9, 2.4) & & (32.1, 29.5) & (7.0, 4.7) & (4.6, 2.2) & & (37.1, 33.7) & (7.9, 4.8) & (5.5, 2.2) \\
    \cline{2-4} \cline{6-8} \cline{10-12}
        &  \multicolumn{11}{c}{$\gamma_0=0.3$} \\
0.50 & (104.8, 103.0) & (4.3, 2.8) & (2.3, 0.7) & & (64.5, 61.8) & (3.9, 1.5) & (3.1, 0.3) & & (43.8, 40.4) & (4.5, 1.0) & (4.0, 0.2) \\
0.75 & (219.3, 217.5) & (36.5, 34.8) & (15.1, 13.4) & & (176.8, 174.0) & (24.1, 21.6) & (10.9, 8.5) & & (146.6, 143.0) & (19.3, 16.1) & (9.7, 6.6) \\
0.90 & (305.9, 304.0) & (149.8, 148.0) & (96.9, 95.1) & & (282.3, 279.4) & (119.9, 117.1) & (74.8, 72.1) & & (262.7, 259.0) & (102.9, 99.4) & (63.6, 60.1) \\
1.10 & (167.8, 166.0) & (86.2, 84.5) & (61.3, 59.6) & & (176.1, 173.2) & (82.8, 80.1) & (57.1, 54.4) & & (184.3, 180.6) & (82.3, 78.7) & (55.8, 52.3) \\
1.25 & (71.7, 69.9) & (21.9, 20.3) & (13.0, 11.4) & & (79.4, 76.7) & (21.8, 19.3) & (12.9, 10.5) & & (87.1, 83.6) & (22.5, 19.2) & (13.4, 10.3) \\
1.50 & (29.4, 27.7) & (6.9, 5.4) & (4.2, 2.7) & & (34.5, 32.0) & (7.7, 5.3) & (5.0, 2.6) & & (39.8, 36.4) & (8.6, 5.5) & (5.8, 2.6) \\
    \cline{2-4} \cline{6-8} \cline{10-12}
        &  \multicolumn{11}{c}{$\gamma_0=0.4$} \\
0.50 & (107.7, 105.8) & (4.6, 3.1) & (2.4, 0.8) & & (66.9, 64.2) & (4.1, 1.7) & (3.1, 0.4) & & (45.7, 42.2) & (4.5, 1.1) & (4.0, 0.2) \\
0.75 & (222.9, 221.0) & (39.4, 37.6) & (16.5, 14.9) & & (180.8, 178.0) & (26.1, 23.6) & (11.9, 9.5) & & (150.7, 147.1) & (20.9, 17.6) & (10.5, 7.3) \\
0.90 & (308.0, 306.1) & (155.5, 153.7) & (102.5, 100.7) & & (285.0, 282.2) & (125.4, 122.7) & (79.8, 77.1) & & (266.0, 262.3) & (108.3, 104.7) & (68.1, 64.6) \\
1.10 & (173.9, 172.0) & (92.6, 90.8) & (66.9, 65.2) & & (181.7, 178.9) & (88.8, 86.1) & (62.2, 59.6) & & (189.8, 186.1) & (88.1, 84.5) & (60.7, 57.2) \\
1.25 & (77.2, 75.4) & (24.8, 23.2) & (14.9, 13.3) & & (84.8, 82.1) & (24.4, 21.9) & (14.6, 12.1) & & (92.6, 89.1) & (25.0, 21.7) & (15.0, 11.8) \\
1.50 & (33.0, 31.3) & (8.0, 6.5) & (4.8, 3.3) & & (38.3, 35.7) & (8.7, 6.3) & (5.5, 3.1) & & (43.8, 40.4) & (9.6, 6.4) & (6.3, 3.1) \\
    \cline{2-4} \cline{6-8} \cline{10-12}
        &  \multicolumn{11}{c}{$\gamma_0=0.5$} \\
0.50 & (111.3, 109.5) & (5.0, 3.5) & (2.5, 0.9) & & (69.9, 67.3) & (4.3, 1.9) & (3.2, 0.5) & & (48.1, 44.7) & (4.7, 1.3) & (4.1, 0.3) \\
0.75 & (227.3, 225.4) & (42.9, 41.2) & (18.4, 16.8) & & (185.8, 183.0) & (28.7, 26.1) & (13.2, 10.8) & & (155.9, 152.3) & (22.9, 19.6) & (11.5, 8.3) \\
0.90 & (310.6, 308.8) & (162.2, 160.4) & (109.1, 107.3) & & (288.5, 285.6) & (132.1, 129.3) & (85.8, 83.1) & & (270.0, 266.3) & (114.7, 111.1) & (73.6, 70.1) \\
1.10 & (181.9, 180.0) & (100.7, 98.9) & (74.0, 72.3) & & (189.1, 186.3) & (96.3, 93.6) & (68.7, 66.0) & & (196.7, 193.0) & (95.3, 91.7) & (66.8, 63.3) \\
1.25 & (84.8, 83.0) & (28.8, 27.1) & (17.5, 15.9) & & (92.1, 89.4) & (27.9, 25.4) & (16.8, 14.3) & & (99.9, 96.3) & (28.4, 25.1) & (17.1, 13.9) \\
1.50 & (38.3, 36.5) & (9.5, 8.0) & (5.6, 4.1) & & (43.6, 41.0) & (10.0, 7.7) & (6.2, 3.9) & & (49.3, 45.9) & (10.9, 7.8) & (7.0, 3.8) \\
\hline
    \end{tabular}}
    \vspace{0.2cm}
    \caption{ The values of $ARL_1$ and $SDRL_1$ using the RR$_{r,s}-$MCV control charts for \textbf{$p=4$} and different values of  $\tau,n$ and $\gamma_0$.}
  \label{tab:tabARL3}
  \end{table}

\begin{itemize}
\item The in-control value $\gamma_0$ of MCV and the value of multivariate level $p$ have strong influence on the performance of RR$_{r,s}-$MCV control charts. In particular, the increase of $\gamma_0$ and $p$ results in the increase of the $ARL_1$. For example, in Table \ref{tab:tabARL1} with $\tau=0.5, n=5$ and $(r,s)=(2,3)$, we have $ARL_1=14.2$ when $\gamma_0=0.1$ and $ARL_1=17.1$ when $\gamma_0=0.5$. Also, with the same $\tau=0.5,n=5,\gamma_0=0.2$ and $(r,s)=(2,3)$, we have $ARL_1=14.6$ when $p=2$ in Table \ref{tab:tabARL1} and $ARL_1=102.7$ when $p=4$ in Table \ref{tab:tabARL3}. That is to say, the RR$_{r,s}-$MCV charts are more efficient for processes with small values of in-control MCV and multivariate levels.
\item The sample size $n$ has positive impact on the power of proposed charts: the larger the sample size, the smaller the average number of samples needed to detect the out-of-control status. For instance, with $\gamma_0=0.3, \tau=1.25$ and $p=3$ (Table \ref{tab:tabARL2}), we have $ARL_1=50.1$ when $n=5$ but this value dropped significantly to $ARL_1=11.9$ when $n=15$ on RR$_{3,4}-$MCV chart.
\item Larger values for $(r,s)$ do not necessarily deliver better performance for the Run Rules based control charts; it depends on the value of the sample size $n$, the shift size $\tau$ and especially the type of control chart. In general, using larger values for $(r,s)$ results in better performance for the lower-sided chart  but worse performance for upper-sided chart. For example, with $n=5,\gamma_0=0.5$ in Table \ref{tab:tabARL3}, the RR$_{2,3}-$MCV charts results in $ARL_1=227.3$ for $\tau=0.75$ (lower-sided) and $ARL_1=181.9$ for $\tau=1.1$ (upper-sided) compared to $ARL_1=155.9$ for $\tau=0.75$ (lower-sided) and $ARL_1=196.7$ for $\tau=1.1$ (upper-sided) in the RR$_{4,5}-$MCV charts. 
\end{itemize}

To compare the performance of the RR$_{r,s}-$MCV control charts with the Shewhart$-$ MCV control chart, we define the index $\Delta_A$ as
\begin{equation}
  \Delta_A=100\times\frac{ARL_{\mathrm{Shewhart}}-ARL_{\mathrm{RR}_{r,s}}}{ARL_{\mathrm{RR}_{r,s}}}.
\end{equation} 
In this definition, $ARL_{\mathrm{Shewhart}}$ and $ARL_{\mathrm{RR}_{r,s}}$  represent the $ARL$ value of the Shewhart-MCV  chart and RR$_{r,s}-\mathrm{MCV}$ chart, respectively. Values $\Delta_A>0$ indicate that the RR$_{r,s}-$MCV charts outperform the Shewhart-MCV chart; conversely, values $\Delta_A<0$ indicate that the Shewhart-MCV chart outperforms the RR$_{r,s}$-MCV charts. Tables \ref{tab:tabdeltaARL_1}-\ref{tab:tabdeltaARL_3} present the rounded results (to the nearest integer)  of $\Delta_A$. It can be seen from these tables that the RR$_{r,s}$-MCV charts outperforms the Shewhart-MCV chart in most cases. 

\begin{figure}
  \hspace{-10mm}
\subfloat[$p=2, n=5, \gamma_0=0.1$]
  {\includegraphics[width=.6\linewidth]{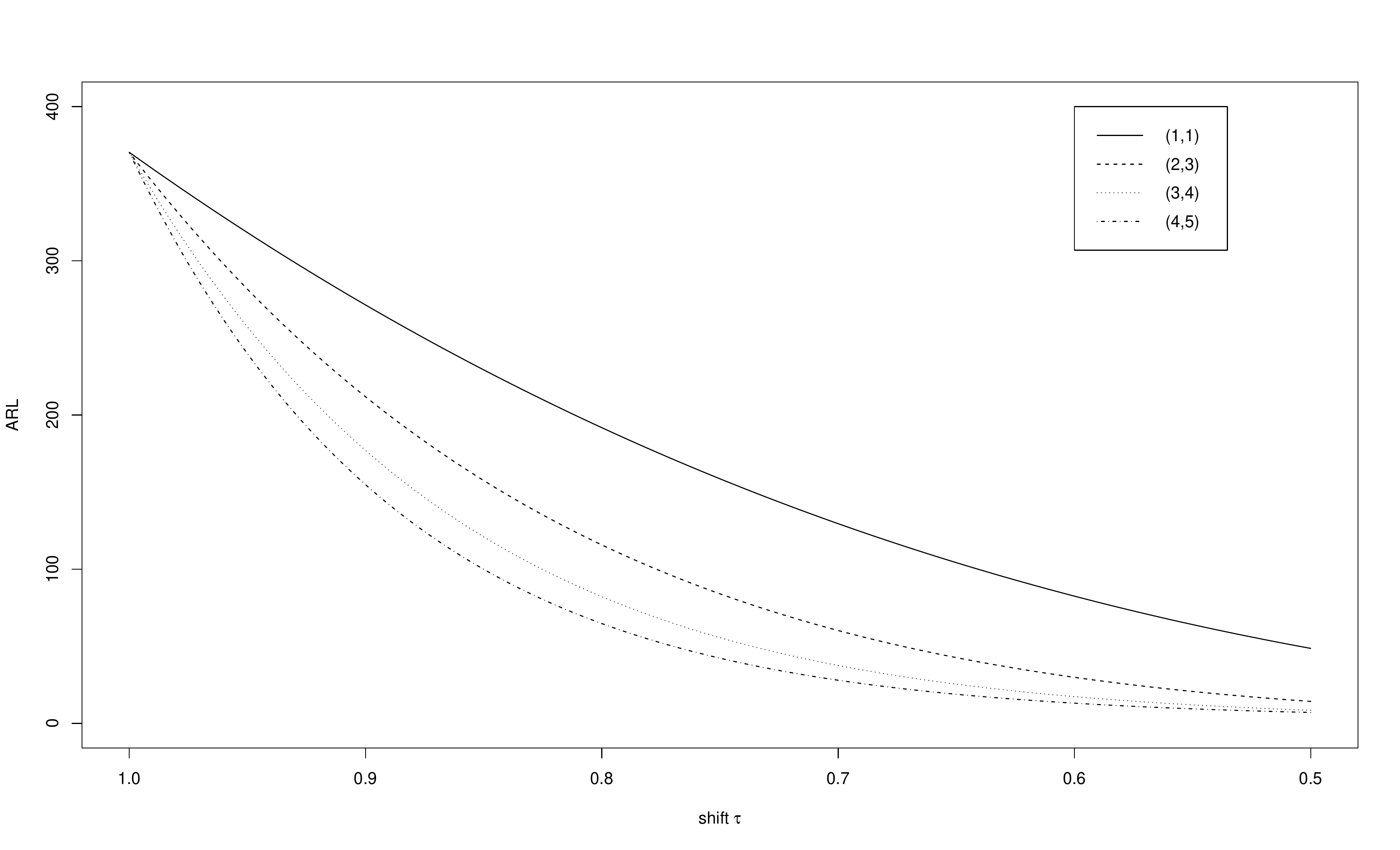}}\hfill
\subfloat[$p=2, n=5, \gamma_0=0.1$]
  {\includegraphics[width=.6\linewidth]{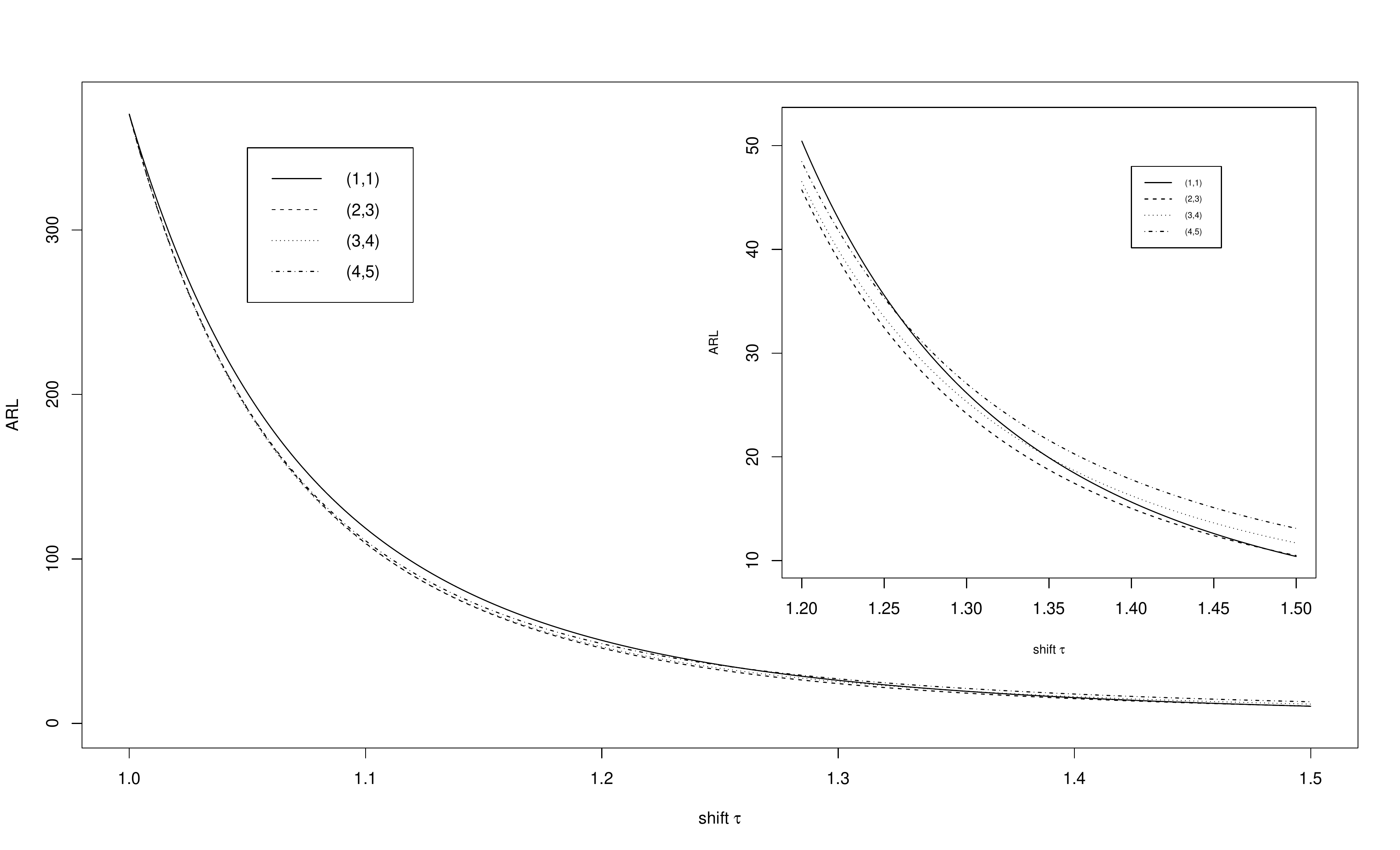}}\hfill

\hspace{-10mm}
\subfloat[$ p=3, n=10, \gamma_0=0.3$]
  {\includegraphics[width=.6\linewidth]{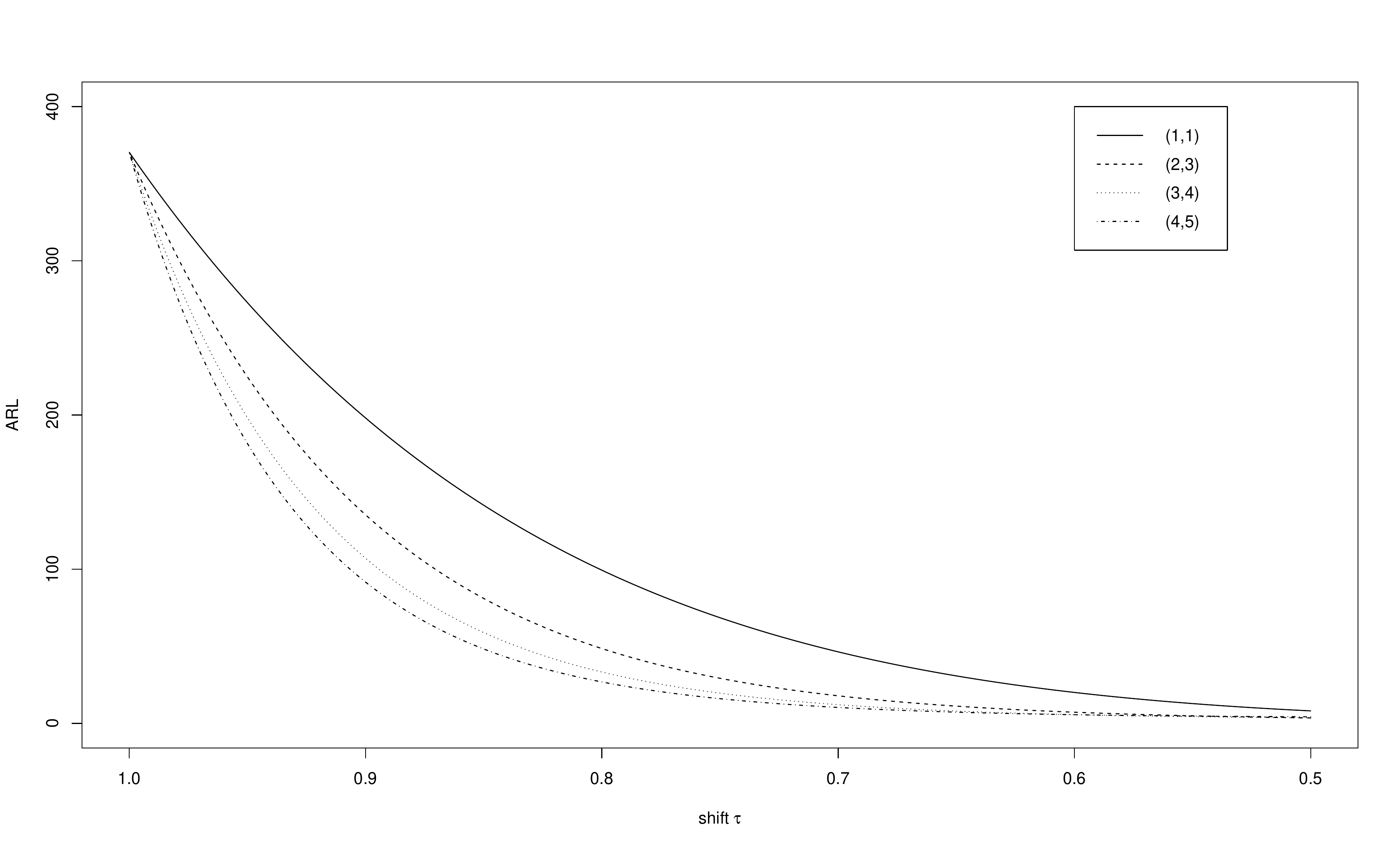}}\hfill
\subfloat[$ p=3, n=10, \gamma_0=0.3$]
  {\includegraphics[width=.6\linewidth]{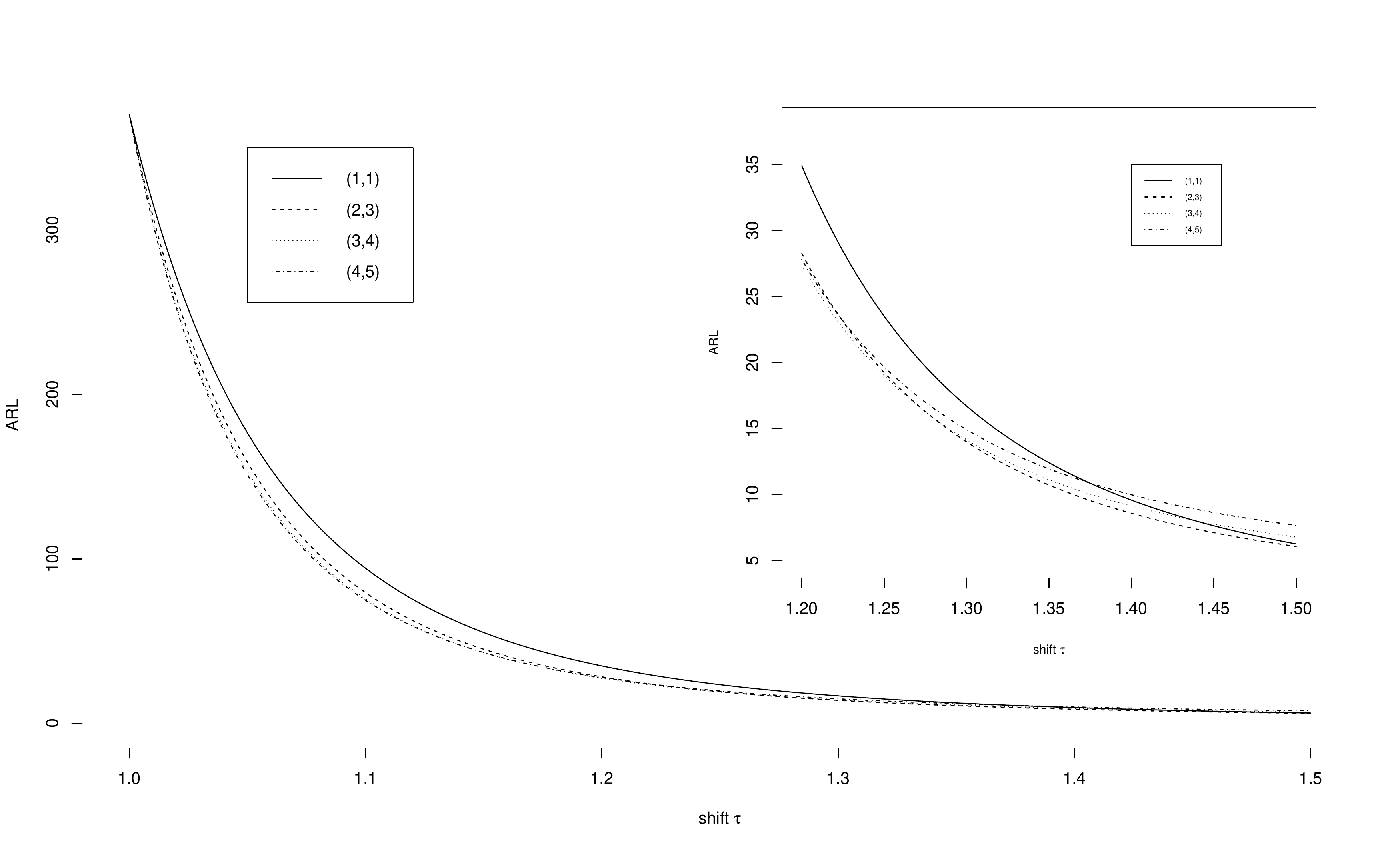}}\hfill
  
\hspace{-10mm}
\subfloat[ $p=4, n=15, \gamma_0=0.5$]
  {\includegraphics[width=.6\linewidth]{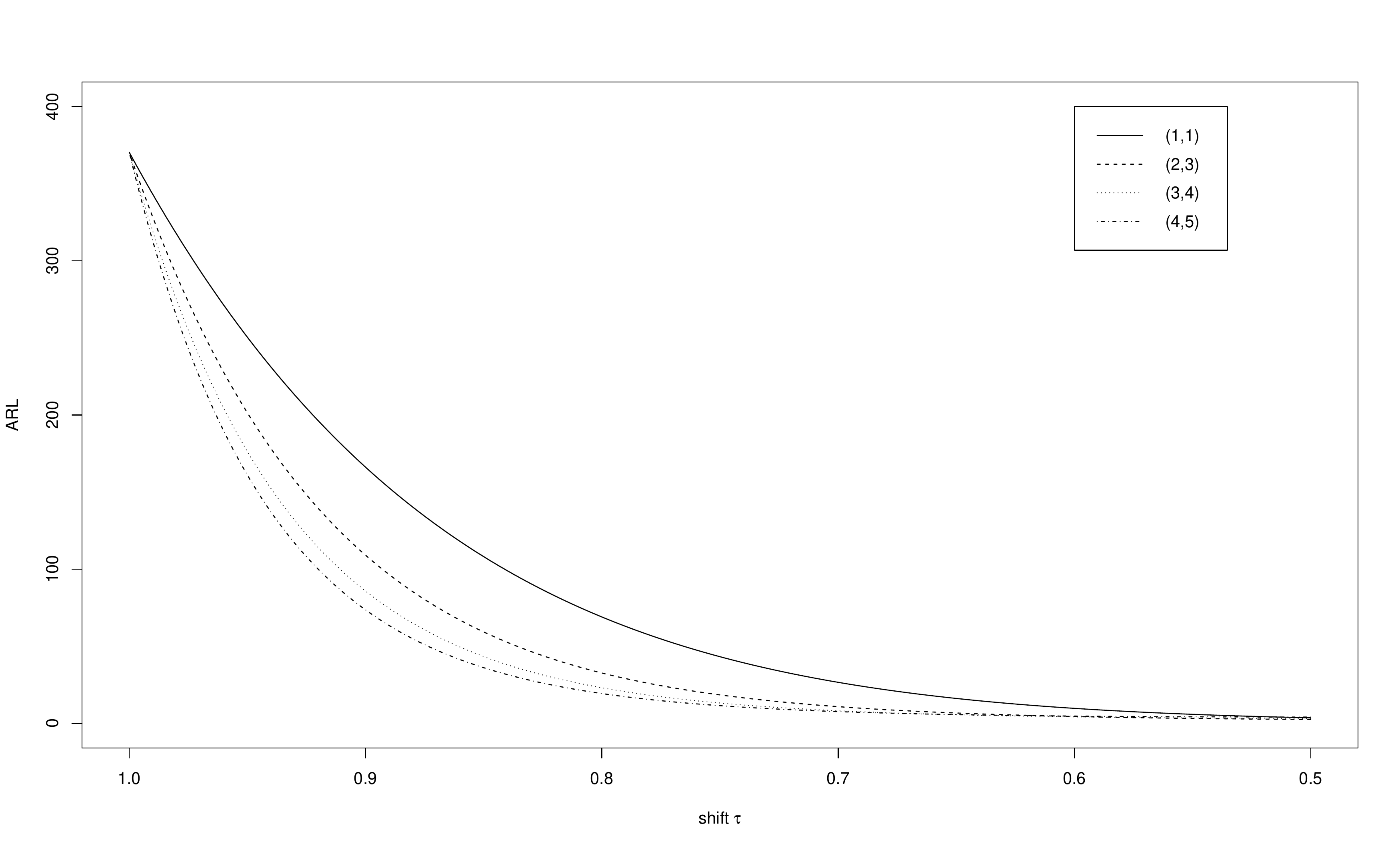}}\hfill
\subfloat[$ p=4, n=15, \gamma_0=0.5$]
  {\includegraphics[width=.6\linewidth]{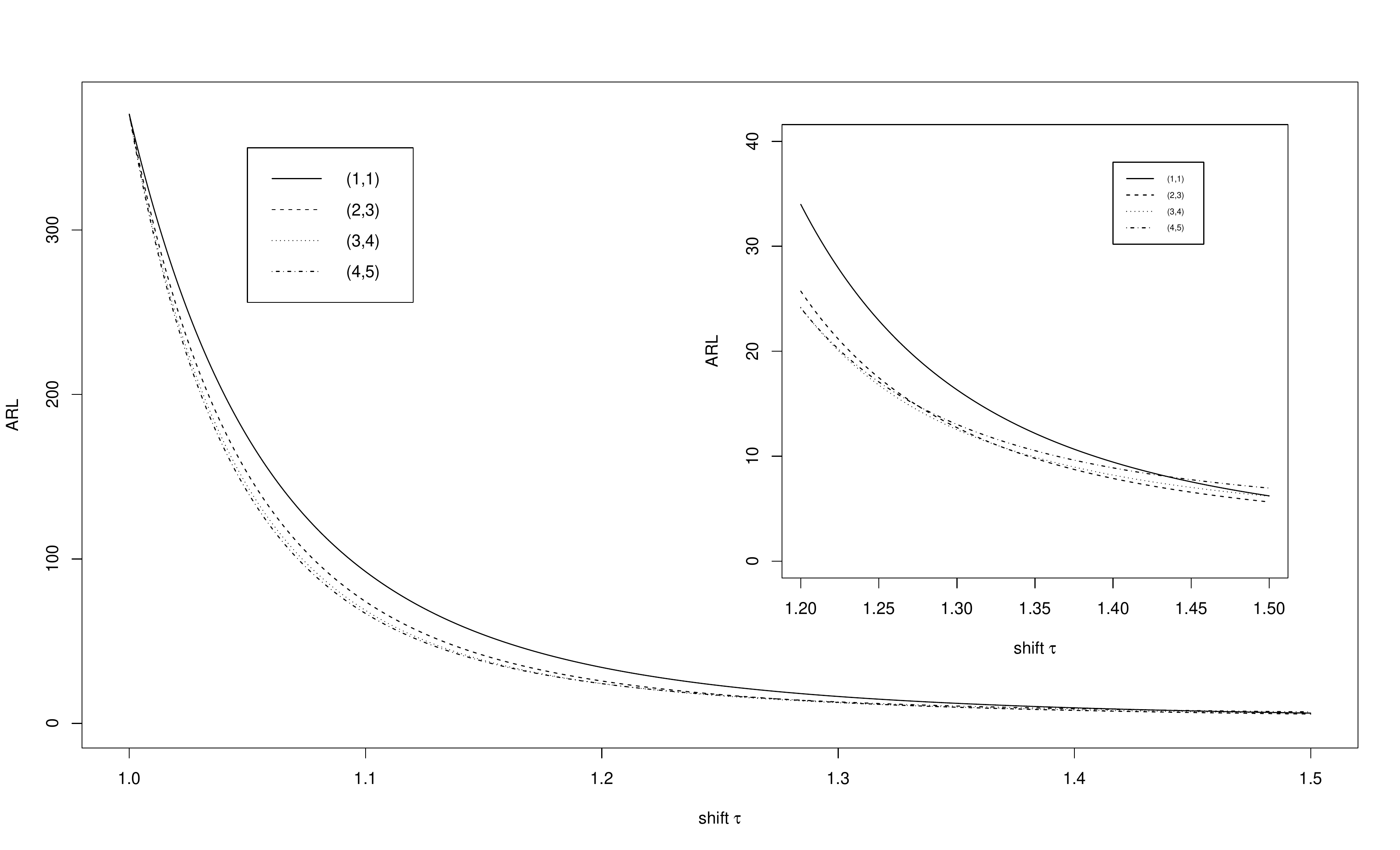}}\hfill

\caption{The $ARL$ profiles of Shewhart chart and Run Rules charts for various in-control settings; left side: lower-sided charts, right side: upper-sided charts \label{graph}}
\end{figure}

\vspace{0.25cm}
The above conclusions can be seen more clearly in Figure \ref{graph}, where we present the $ARL$ profiles for both the Shewhart chart (designed by \citet{Yeong_MCV_2016}) and the Run Rules charts for a number of different in-control scenarios. Since the $ARL$ curves for the upper case overlap each other, we include the ARL curves for $\tau\geqslant 1.2$ as an inset plot. The Figure 1 shows that for the lower case (decrease shifts), the 4-of-5 Run Rule chart remarkably outperforms the Shewhart chart and the other Run Rules charts, especially when $n=5,p=2$ and $\gamma_0=0.1$. As $n, p, \gamma_0$ increase, the improvement is not as much as in the first case but still, it is substantial. For upper case (increase shifts), we have also an improvement with Run Rules charts but it is not as much as in the lower case. In addition, a part of the $ARL$ curve of the Shehwart chart corresponding to very large  shifts (i.e., 1.50 or  0.50) is below $ARL$ curves of Run Rules charts. We deduce that the Shewhart chart becomes more efficient than the proposed Run Rule based charts in detecting very large shifts. This is also confirmed by the negative values of $\Delta_A$ index with the large values of $\tau$ in the Tables \ref{tab:tabdeltaARL_1}-\ref{tab:tabdeltaARL_3}.

\begin{table}[thb]
  \scalebox{0.9}{
    \begin{tabular}{cccccccccccc}
     \hline
  & \multicolumn{3}{c}{RR$_{2,3}$-MCV chart} & &\multicolumn{3}{c}{RR$_{3,4}$-MCV chart} && \multicolumn{3}{c}{RR$_{4,5}$-MCV chart}\\
      \cline{2-4} \cline{6-8} \cline{10-12}
$\tau$&  $n=5$ & $n=10$ & $n=15$  &&  $n=5$ & $n=10$ & $n=15$ &&  $n=5$ & $n=10$ & $n=15$ \\
     \cline{2-4} \cline{6-8} \cline{10-12}
        &  \multicolumn{11}{c}{$\gamma_0=0.1$} \\
0.50 & 71 & 46 & -10 & & 83 & 38 & -56 & & 85 & 22 & -107 \\
0.75 & 47 & 59 & 58 & & 65 & 72 & 68 & & 73 & 76 & 70 \\
0.90 & 22 & 35 & 40 & & 35 & 49 & 55 & & 43 & 57 & 62 \\
1.10 &  8 & 17 & 22 & &  8 & 22 & 28 & &  6 & 23 & 30 \\
1.25 &  9 & 18 & 18 & &  6 & 18 & 17 & &  1 & 14 & 12 \\
1.50 & -1 & -5 & -17 & & -13 & -22 & -45 & & -26 & -42 & -77 \\
     \cline{2-4} \cline{6-8} \cline{10-12}
        &  \multicolumn{11}{c}{$\gamma_0=0.2$} \\
0.50 & 71 & 47 & -7 & & 83 & 40 & -51 & & 85 & 25 & -100 \\
0.75 & 47 & 59 & 58 & & 65 & 72 & 68 & & 73 & 76 & 70 \\
0.90 & 22 & 34 & 39 & & 34 & 49 & 54 & & 42 & 56 & 61 \\
1.10 &  8 & 17 & 22 & &  8 & 22 & 28 & &  6 & 23 & 30 \\
1.25 &  9 & 18 & 19 & &  6 & 18 & 19 & &  1 & 15 & 14 \\
1.50 &  0 & -2 & -13 & & -11 & -18 & -38 & & -24 & -36 & -67 \\
     \cline{2-4} \cline{6-8} \cline{10-12}
        &  \multicolumn{11}{c}{$\gamma_0=0.3$} \\
0.50 & 70 & 49 & -2 & & 82 & 43 & -43 & & 86 & 29 & -88 \\
0.75 & 46 & 58 & 58 & & 64 & 71 & 68 & & 72 & 76 & 71 \\
0.90 & 21 & 33 & 39 & & 34 & 48 & 53 & & 42 & 55 & 60 \\
1.10 &  8 & 17 & 22 & &  8 & 22 & 28 & &  7 & 23 & 30 \\
1.25 & 10 & 19 & 21 & &  7 & 20 & 21 & &  3 & 17 & 17 \\
1.50 &  2 &  2 & -7 & & -7 & -11 & -28 & & -19 & -27 & -53 \\
     \cline{2-4} \cline{6-8} \cline{10-12}
        &  \multicolumn{11}{c}{$\gamma_0=0.4$} \\
0.50 & 70 & 51 &  4 & & 82 & 46 & -32 & & 86 & 35 & -74 \\
0.75 & 45 & 57 & 58 & & 63 & 71 & 68 & & 72 & 76 & 71 \\
0.90 & 21 & 32 & 37 & & 33 & 46 & 52 & & 41 & 54 & 59 \\
1.10 &  8 & 17 & 22 & &  9 & 22 & 28 & &  8 & 23 & 30 \\
1.25 & 11 & 20 & 23 & & 10 & 22 & 24 & &  6 & 20 & 21 \\
1.50 &  6 &  7 & -1 & & -2 & -3 & -17 & & -12 & -15 & -36 \\
     \cline{2-4} \cline{6-8} \cline{10-12}
        &  \multicolumn{11}{c}{$\gamma_0=0.5$} \\
0.50 & 69 & 53 & 11 & & 82 & 50 & -21 & & 85 & 40 & -58 \\
0.75 & 44 & 56 & 57 & & 62 & 70 & 68 & & 71 & 75 & 71 \\
0.90 & 20 & 31 & 36 & & 32 & 45 & 50 & & 39 & 53 & 57 \\
1.10 &  9 & 17 & 21 & & 10 & 22 & 28 & & 10 & 24 & 30 \\
1.25 & 14 & 22 & 25 & & 14 & 25 & 27 & & 11 & 24 & 26 \\
1.50 & 12 & 12 &  7 & &  7 &  6 & -5 & & -1 & -3 & -20 \\
\hline
    \end{tabular}}
      \caption{The values of index $\Delta_A$ for $p=2$ and different values of  $\tau,\gamma_0$ and $n$.}
  \label{tab:tabdeltaARL_1}
  \end{table}

\begin{table}[thb]
  \scalebox{0.9}{
    \begin{tabular}{cccccccccccc}
     \hline
  & \multicolumn{3}{c}{RR$_{2,3}$-MCV chart} & &\multicolumn{3}{c}{RR$_{3,4}$-MCV chart} && \multicolumn{3}{c}{RR$_{4,5}$-MCV chart}\\
      \cline{2-4} \cline{6-8} \cline{10-12}
$\tau$&  $n=5$ & $n=10$ & $n=15$  &&  $n=5$ & $n=10$ & $n=15$ &&  $n=5$ & $n=10$ & $n=15$ \\
     \cline{2-4} \cline{6-8} \cline{10-12}
        &  \multicolumn{11}{c}{$\gamma_0=0.1$} \\
0.50 & 65 & 55 &  2 & & 81 & 52 & -37 & & 87 & 42 & -80 \\
0.75 & 38 & 58 & 59 & & 57 & 72 & 69 & & 67 & 77 & 71 \\
0.90 & 17 & 33 & 39 & & 28 & 47 & 54 & & 36 & 55 & 61 \\
1.10 &  4 & 16 & 21 & &  2 & 20 & 27 & & -1 & 21 & 29 \\
1.25 &  3 & 17 & 19 & & -2 & 17 & 18 & & -9 & 13 & 13 \\
1.50 & -4 & -3 & -14 & & -18 & -18 & -40 & & -33 & -36 & -69 \\
     \cline{2-4} \cline{6-8} \cline{10-12}
        &  \multicolumn{11}{c}{$\gamma_0=0.2$} \\
0.50 & 65 & 56 &  5 & & 81 & 54 & -32 & & 87 & 44 & -73 \\
0.75 & 38 & 58 & 58 & & 56 & 72 & 69 & & 66 & 77 & 72 \\
0.90 & 16 & 33 & 39 & & 27 & 47 & 53 & & 35 & 55 & 60 \\
1.10 &  4 & 16 & 21 & &  2 & 20 & 27 & & -1 & 21 & 29 \\
1.25 &  4 & 17 & 19 & & -2 & 18 & 19 & & -9 & 14 & 15 \\
1.50 & -4 & -1 & -10 & & -17 & -14 & -34 & & -31 & -31 & -60 \\
     \cline{2-4} \cline{6-8} \cline{10-12}
        &  \multicolumn{11}{c}{$\gamma_0=0.3$} \\
0.50 & 64 & 57 &  9 & & 81 & 56 & -24 & & 86 & 47 & -63 \\
0.75 & 37 & 57 & 58 & & 56 & 71 & 69 & & 66 & 77 & 72 \\
0.90 & 16 & 32 & 38 & & 27 & 46 & 52 & & 34 & 54 & 59 \\
1.10 &  4 & 16 & 21 & &  2 & 20 & 27 & & -0 & 21 & 29 \\
1.25 &  4 & 18 & 21 & & -1 & 19 & 21 & & -7 & 16 & 18 \\
1.50 & -2 &  3 & -5 & & -14 & -8 & -24 & & -27 & -22 & -47 \\
     \cline{2-4} \cline{6-8} \cline{10-12}
        &  \multicolumn{11}{c}{$\gamma_0=0.4$} \\
0.50 & 64 & 58 & 15 & & 80 & 58 & -15 & & 86 & 51 & -50 \\
0.75 & 36 & 56 & 58 & & 55 & 71 & 69 & & 65 & 76 & 72 \\
0.90 & 15 & 31 & 37 & & 26 & 45 & 51 & & 33 & 52 & 58 \\
1.10 &  4 & 16 & 21 & &  3 & 20 & 27 & &  1 & 21 & 29 \\
1.25 &  6 & 19 & 22 & &  2 & 21 & 24 & & -4 & 19 & 22 \\
1.50 &  1 &  8 &  1 & & -8 & -1 & -13 & & -20 & -12 & -31 \\
     \cline{2-4} \cline{6-8} \cline{10-12}
        &  \multicolumn{11}{c}{$\gamma_0=0.5$} \\
0.50 & 63 & 59 & 20 & & 80 & 61 & -5 & & 86 & 55 & -36 \\
0.75 & 35 & 55 & 57 & & 53 & 70 & 69 & & 63 & 76 & 73 \\
0.90 & 15 & 30 & 35 & & 25 & 43 & 49 & & 32 & 51 & 57 \\
1.10 &  6 & 16 & 21 & &  5 & 20 & 27 & &  3 & 21 & 29 \\
1.25 &  8 & 21 & 24 & &  6 & 24 & 27 & &  1 & 23 & 26 \\
1.50 &  7 & 13 &  8 & & -0 &  8 & -2 & & -10 & -1 & -16 \\
\hline
    \end{tabular}}
    \caption{ The values of  $\Delta_A$ index for $p=3$ and different values of  $\tau,\gamma_0$ and $n$.}
  \label{tab:tabdeltaARL_2}
  \end{table}

\begin{table}[thb]
  \scalebox{0.9}{
    \begin{tabular}{cccccccccccc}
     \hline
  & \multicolumn{3}{c}{RR$_{2,3}$-MCV chart} & &\multicolumn{3}{c}{RR$_{3,4}$-MCV chart} && \multicolumn{3}{c}{RR$_{4,5}$-MCV chart}\\
      \cline{2-4} \cline{6-8} \cline{10-12}
$\tau$&  $n=5$ & $n=10$ & $n=15$  &&  $n=5$ & $n=10$ & $n=15$ &&  $n=5$ & $n=10$ & $n=15$ \\
     \cline{2-4} \cline{6-8} \cline{10-12}
        &  \multicolumn{11}{c}{$\gamma_0=0.1$} \\
        
0.50 & 45 & 62 & 14 & & 67 & 64 & -17 & & 78 & 59 & -53 \\
0.75 & 23 & 57 & 59 & & 38 & 72 & 70 & & 49 & 77 & 73 \\
0.90 &  9 & 31 & 38 & & 16 & 45 & 53 & & 22 & 53 & 60 \\
1.10 & -2 & 14 & 21 & & -8 & 18 & 26 & & -13 & 18 & 27 \\
1.25 & -7 & 16 & 19 & & -19 & 16 & 18 & & -31 & 12 & 13 \\
1.50 & -15 & -1 & -11 & & -37 & -15 & -35 & & -59 & -31 & -62 \\
     \cline{2-4} \cline{6-8} \cline{10-12}
        &  \multicolumn{11}{c}{$\gamma_0=0.2$} \\
0.50 & 45 & 62 & 17 & & 66 & 65 & -12 & & 77 & 60 & -47 \\
0.75 & 22 & 57 & 59 & & 38 & 72 & 70 & & 49 & 77 & 73 \\
0.90 &  9 & 31 & 38 & & 16 & 45 & 52 & & 22 & 53 & 60 \\
1.10 & -2 & 14 & 20 & & -8 & 18 & 26 & & -13 & 18 & 27 \\
1.25 & -7 & 16 & 19 & & -19 & 16 & 19 & & -31 & 13 & 15 \\
1.50 & -15 &  1 & -8 & & -36 & -12 & -29 & & -57 & -26 & -53 \\
     \cline{2-4} \cline{6-8} \cline{10-12}
        &  \multicolumn{11}{c}{$\gamma_0=0.3$} \\
0.50 & 45 & 63 & 20 & & 66 & 66 & -6 & & 77 & 62 & -38 \\
0.75 & 22 & 56 & 58 & & 37 & 71 & 70 & & 48 & 77 & 73 \\
0.90 &  9 & 30 & 37 & & 16 & 44 & 51 & & 22 & 52 & 59 \\
1.10 & -2 & 14 & 20 & & -7 & 18 & 26 & & -12 & 18 & 27 \\
1.25 & -6 & 17 & 20 & & -17 & 18 & 21 & & -29 & 15 & 18 \\
1.50 & -13 &  4 & -3 & & -33 & -6 & -21 & & -53 & -19 & -41 \\
     \cline{2-4} \cline{6-8} \cline{10-12}
        &  \multicolumn{11}{c}{$\gamma_0=0.4$} \\
0.50 & 44 & 64 & 25 & & 65 & 68 &  3 & & 76 & 64 & -26 \\
0.75 & 21 & 55 & 58 & & 36 & 70 & 70 & & 47 & 76 & 73 \\
0.90 &  8 & 29 & 36 & & 15 & 43 & 50 & & 21 & 50 & 57 \\
1.10 & -1 & 14 & 20 & & -6 & 18 & 26 & & -11 & 18 & 27 \\
1.25 & -4 & 18 & 22 & & -15 & 20 & 24 & & -25 & 17 & 22 \\
1.50 & -10 &  8 &  3 & & -28 &  1 & -10 & & -46 & -10 & -27 \\
     \cline{2-4} \cline{6-8} \cline{10-12}
        &  \multicolumn{11}{c}{$\gamma_0=0.5$} \\
0.50 & 43 & 64 & 30 & & 64 & 69 & 11 & & 75 & 66 & -14 \\
0.75 & 21 & 54 & 57 & & 35 & 69 & 70 & & 46 & 75 & 73 \\
0.90 &  8 & 28 & 34 & & 15 & 41 & 48 & & 20 & 49 & 56 \\
1.10 & -0 & 14 & 20 & & -4 & 18 & 26 & & -9 & 19 & 28 \\
1.25 & -2 & 20 & 24 & & -11 & 22 & 27 & & -20 & 21 & 26 \\
1.50 & -5 & 13 &  9 & & -20 &  9 &  0 & & -35 &  1 & -12 \\
\hline
    \end{tabular}}
      \caption{The values of $\Delta_A$ index for $p=4$ and different values of  $\tau,\gamma_0$ and $n$.}
  \label{tab:tabdeltaARL_3}
  \end{table}

  \vspace{0.25cm}
The analysis presented above is only for the case of specific shift size. In practice, however,  it is hard  for quality practitioners to predetermine a specific shift without any previous experience. Thus, they usually have an interest in detecting a range of shifts $\tau\in[a,b]$ rather than preference for any particular size of the process shift. The use of the uniform distribution has been proposed to account for the unknown shift size by some authors (for instance, see \citet{Chen2007b} and \citet{Celano2014a}). The statistical performance of the corresponding chart can be
evaluated through the $EARL$ (Expected Average Run Length) given by
\begin{equation}
  \label{equ:EARL}
  EARL=\int_a^bARL(\tau)\times f_{\tau}(\tau)d\tau,
\end{equation}
\noindent
where $ARL(\tau)$ is the $ARL$, as a function of shift $\tau$, of the proposed Run Rules charts, with  $f_{\tau}(\tau)=\frac{1}{b-a}$ for $\tau \in [a,b]$. In the following section, we will consider a specific 
range of shift $[a,b]=[0.5,1)$ (decreasing case, denoted as (D)) for lower-sided RR$_{r,s}-$MCV control chart and $[a,b]=(1,2]$ (increasing case, denoted as (I))  for upper-sided RR$_{r,s}-$MCV control chart. \\

\begin{table}[thb]
\hspace*{-10mm}
  \scalebox{0.675}{
    \begin{tabular}{cccccccccccc}
     \hline
       &  \multicolumn{11}{c}{$p=2$} \\
      \hline
  & \multicolumn{3}{c}{RR$_{2,3}$-MCV chart} & &\multicolumn{3}{c}{RR$_{3,4}$-MCV chart} && \multicolumn{3}{c}{RR$_{4,5}$-MCV chart}\\
  &  \multicolumn{11}{c}{$p=2$} \\
      \cline{2-4} \cline{6-8} \cline{10-12}
$\tau$&  $n=5$ & $n=10$ & $n=15$  &&  $n=5$ & $n=10$ & $n=15$ &&  $n=5$ & $n=10$ & $n=15$ \\
     \cline{2-4} \cline{6-8} \cline{10-12}
        &  \multicolumn{11}{c}{$\gamma_0=0.1$} \\
        
(D) & (101.8, 100.1) & (48.1, 46.5) & (33.0, 31.3) & & (79.4,76.7) & (38.5,36.0) & (26.9,24.3) & & (67.8,64.4) & (34.0,30.6) & (24.2,20.7) \\
 (I) & (29.4, 27.8) & (17.4, 15.8) & (13.2, 11.5) & & (30.3,27.9) & (17.4,15.0) & (13.1,10.6) & & (31.7,28.5) & (18.0,14.7) & (13.7,10.2) \\

     \cline{2-4} \cline{6-8} \cline{10-12}
        &  \multicolumn{11}{c}{$\gamma_0=0.2$} \\
 (D) & (103.1, 101.3) & (49.2, 47.5) & (33.9, 32.2) & & (80.4,77.8) & (39.4,36.9) & (27.6,25.1) & & (68.8,65.4) & (34.8,31.4) & (24.9,21.4) \\
 (I) & (30.3, 28.6) & (18.1, 16.5) & (13.7, 12.1) & & (31.2,28.7) & (18.1,15.6) & (13.6,11.1) & & (32.6,29.3) & (18.6,15.3) & (14.1,10.7) \\
     \cline{2-4} \cline{6-8} \cline{10-12}
        &  \multicolumn{11}{c}{$\gamma_0=0.3$} \\
 (D) & (105.1, 103.3) & (50.9, 49.2) & (35.3, 33.7) & & (82.2,79.6) & (40.9,38.4) & (28.8,26.3) & & (70.4,67.0) & (36.1,32.7) & (26.0,22.5) \\
 (I) & (31.8, 30.1) & (19.2, 17.7) & (14.6, 13.0) & & (32.6,30.1) & (19.1,16.7) & (14.5,12.0) & & (34.0,30.8) & (19.6,16.3) & (14.9,11.5) \\
     \cline{2-4} \cline{6-8} \cline{10-12}
        &  \multicolumn{11}{c}{$\gamma_0=0.4$} \\
(D) & (107.8, 106.1) & (53.2, 51.5) & (37.2, 35.6) & & (84.7,82.0) & (42.9,40.3) & (30.5,27.9) & & (72.6,69.2) & (37.9,34.5) & (27.4,24.0) \\
 (I) & (34.0, 32.4) & (20.9, 19.3) & (15.9, 14.3) & & (34.7,32.3) & (20.6,18.1) & (15.6,13.1) & & (36.1,32.9) & (21.1,17.8) & (16.0,12.6) \\
     \cline{2-4} \cline{6-8} \cline{10-12}
        &  \multicolumn{11}{c}{$\gamma_0=0.5$} \\
 (D) & (111.2, 109.4) & (55.9, 54.3) & (39.5, 37.9) & & (87.7,85.0) & (45.2,42.7) & (32.4,29.9) & & (75.4,72.0) & (40.1,36.7) & (29.2,25.8) \\
 (I) & (37.2, 35.5) & (23.0, 21.4) & (17.5, 16.0) & & (37.6,35.1) & (22.5,20.1) & (17.1,14.6) & & (38.9,35.6) & (22.9,19.6) & (17.4,14.1) \\
      \hline
       &  \multicolumn{11}{c}{$p=3$} \\
      \hline
  & \multicolumn{3}{c}{RR$_{2,3}$-MCV chart} & &\multicolumn{3}{c}{RR$_{3,4}$-MCV chart} && \multicolumn{3}{c}{RR$_{4,5}$-MCV chart}\\
      \cline{2-4} \cline{6-8} \cline{10-12}
$\tau$&  $n=5$ & $n=10$ & $n=15$  &&  $n=5$ & $n=10$ & $n=15$ &&  $n=5$ & $n=10$ & $n=15$ \\
     \cline{2-4} \cline{6-8} \cline{10-12}
        &  \multicolumn{11}{c}{$\gamma_0=0.1$} \\
(D) & (136.9, 135.1) & (53.3, 51.7) & (35.1, 33.5) & & (107.0,104.3) & (42.5,40.0) & (28.5,26.0) & & (90.6,87.2) & (37.3,34.0) & (25.6,22.1) \\
 (I) & (36.0, 34.3) & (18.8, 17.2) & (13.8, 12.2) & & (37.9,35.4) & (18.8,16.4) & (13.8,11.2) & & (40.2,36.9) & (19.5,16.2) & (14.3,10.9) \\
     \cline{2-4} \cline{6-8} \cline{10-12}
        &  \multicolumn{11}{c}{$\gamma_0=0.2$} \\
 (D) & (138.3, 136.5) & (54.5, 52.8) & (36.1, 34.4) & & (108.3,105.6) & (43.5,41.0) & (29.3,26.8) & & (91.8,88.3) & (38.2,34.8) & (26.3,22.9) \\
 (I) & (37.0, 35.4) & (19.5, 17.9) & (14.4, 12.8) & & (38.9,36.4) & (19.5,17.1) & (14.3,11.8) & & (41.2,38.0) & (20.1,16.8) & (14.8,11.4) \\
     \cline{2-4} \cline{6-8} \cline{10-12}
        &  \multicolumn{11}{c}{$\gamma_0=0.3$} \\
 (D) & (140.6, 138.8) & (56.3, 54.7) & (37.6, 35.9) & & (110.3,107.6) & (45.1,42.5) & (30.6,28.0) & & (93.7,90.2) & (39.6,36.3) & (27.4,24.0) \\
 (I) & (38.8, 37.2) & (20.7, 19.2) & (15.3, 13.7) & & (40.7,38.2) & (20.7,18.2) & (15.2,12.7) & & (43.1,39.8) & (21.2,18.0) & (15.6,12.3) \\
     \cline{2-4} \cline{6-8} \cline{10-12}
        &  \multicolumn{11}{c}{$\gamma_0=0.4$} \\
 (D) & (143.6, 141.8) & (58.8, 57.1) & (39.6, 37.9) & & (113.2,110.5) & (47.2,44.7) & (32.3,29.8) & & (96.4,92.9) & (41.5,38.2) & (29.0,25.6) \\
 (I) & (41.6, 40.0) & (22.5, 20.9) & (16.7, 15.1) & & (43.4,40.9) & (22.3,19.8) & (16.4,13.9) & & (45.8,42.5) & (22.8,19.6) & (16.8,13.4) \\

     \cline{2-4} \cline{6-8} \cline{10-12}
        &  \multicolumn{11}{c}{$\gamma_0=0.5$} \\
 (D) & (147.5, 145.7) & (61.8, 60.1) & (42.0, 40.4) & & (116.8,114.1) & (49.8,47.2) & (34.4,31.8) & & (99.7,96.2) & (43.9,40.5) & (30.9,27.4) \\
 (I) & (45.6, 43.9) & (24.8, 23.3) & (18.4, 16.9) & & (47.2,44.6) & (24.4,22.0) & (18.0,15.5) & & (49.5,46.2) & (24.8,21.6) & (18.3,15.0) \\

           \hline
       &  \multicolumn{11}{c}{$p=4$} \\
      \hline
  & \multicolumn{3}{c}{RR$_{2,3}$-MCV chart} & &\multicolumn{3}{c}{RR$_{3,4}$-MCV chart} && \multicolumn{3}{c}{RR$_{4,5}$-MCV chart}\\
      \cline{2-4} \cline{6-8} \cline{10-12}
$\tau$&  $n=5$ & $n=10$ & $n=15$  &&  $n=5$ & $n=10$ & $n=15$ &&  $n=5$ & $n=10$ & $n=15$ \\
     \cline{2-4} \cline{6-8} \cline{10-12}

        &  \multicolumn{11}{c}{$\gamma_0=0.1$} \\
(D) & (208.8, 206.9) & (60.1, 58.4) & (37.6, 35.9) & & (171.7,168.9) & (47.6,45.1) & (30.4,27.9) & & (147.1,143.5) & (41.6,38.2) & (27.2,23.8) \\
 (I) & (50.0, 48.3) & (20.5, 18.9) & (14.5, 12.9) & & (55.1,52.6) & (20.6,18.1) & (14.5,12.0) & & (60.3,57.0) & (21.3,18.0) & (15.0,11.6) \\

  &  \multicolumn{11}{c}{$\gamma_0=0.2$} \\
(D) & (210.1, 208.2) & (61.3, 59.6) & (38.6, 36.9) & & (173.1,170.3) & (48.7,46.2) & (31.3,28.7) & & (148.5,144.9) & (42.5,39.2) & (27.9,24.5) \\
 (I) & (51.4, 49.7) & (21.3, 19.7) & (15.1, 13.5) & & (56.6,54.1) & (21.3,18.9) & (15.0,12.5) & & (61.9,58.5) & (22.0,18.7) & (15.6,12.2) \\
  &  \multicolumn{11}{c}{$\gamma_0=0.3$} \\
  (D) & (212.3, 210.4) & (63.3, 61.6) & (40.2, 38.5) & & (175.4,172.6) & (50.4,47.9) & (32.6,30.1) & & (150.8,147.2) & (44.1,40.7) & (29.2,25.7) \\
 (I) & (53.9, 52.2) & (22.6, 21.0) & (16.1, 14.5) & & (59.2,56.7) & (22.6,20.2) & (16.0,13.5) & & (64.7,61.2) & (23.3,20.0) & (16.4,13.1) \\
  &  \multicolumn{11}{c}{$\gamma_0=0.4$} \\
(D) & (215.2, 213.4) & (66.0, 64.3) & (42.3, 40.6) & & (178.6,175.8) & (52.7,50.2) & (34.4,31.9) & & (154.0,150.4) & (46.2,42.8) & (30.8,27.4) \\
 (I) & (57.7, 56.0) & (24.5, 22.9) & (17.6, 16.0) & & (63.1,60.5) & (24.4,22.0) & (17.3,14.8) & & (68.7,65.3) & (25.0,21.8) & (17.7,14.4) \\
  &  \multicolumn{11}{c}{$\gamma_0=0.5$} \\
(D) & (218.9, 217.1) & (69.3, 67.6) & (44.8, 43.2) & & (182.6,179.8) & (55.5,53.0) & (36.6,34.1) & & (158.0,154.4) & (48.8,45.4) & (32.8,29.4) \\
 (I) & (63.1, 61.4) & (27.1, 25.5) & (19.5, 17.9) & & (68.5,65.8) & (26.8, 24.3) & (19.0,16.5) & & (74.1,70.7) & (27.3,24.0) & (19.3,16.0) \\
\hline
    \end{tabular}}
        \caption{The values $(EARL_1,ESDRL_1)$ for $RR_{r,s}^{-}-$MCV control chart when $[a,b]=[0.5,1)$ and for RR$^+_{r,s}-$MCV control chart when $[a,b]=(1,2]$ with different values of $p, n, \gamma_0$ and $\tau$.}
  \label{tab:tabEARL}
  \end{table}

Table \ref{tab:tabEARL} presents the values of $EARL$ and $ESDRL$ (Expected Standard Deviation Run Length) for various combinations of $n=\{5,10,15\}$, $\gamma_0=\{0.1,0.2,0.3,0.4,0.5\}$ and $p=\{2,3,4\}$. The same trends as the case of specific shift size are observed from this table. The value of $EARL$ in the upper-sided Run Rule control chart corresponding to smaller values of $(r,s)$ is significantly smaller than that corresponding to larger values of $(r,s)$
In the contrary, the values of $EARL$ decrease from smaller $(r,s)$ scheme to larger $(r,s)$ scheme of Run Rules for  lower-sided chart. Therefore, the choice of using RR$_{2,3}-$MCV, RR$_{3,4}-$MCV or RR$_{4,5s}-$MCV control charts depends on the goal of practitioners: if they want to detect increasing shifts, they are advised to choose smaller $(r,s)$ scheme of Run Rules (say RR$_{2,3}-$MCV in this paper); otherwise, the larger $(r,s)$ scheme of Run Rules (say RR$_{4,5}-$MCV in this paper) should be used.

\vspace{0.25cm}
Similar to the specific shift size case, a comparison between the performance of RR$_{r,s}$-MCV
control charts and the performance of Shewhart-MCV control chart is provided in
Table \ref{tab:tabDeltaEARL}. It has been undertaken by defining the index
\begin{equation}
  \Delta_E=100\times\frac{EARL_{\mathrm{Shewhart}}-EARL_{\mathrm{RR}_{r,s}}}{EARL_{\mathrm{RR}_{r,s}}},
\end{equation}
where $EARL_{\mathrm{Shewhart}}$ and $EARL_{\mathrm{RR}_{r,s}}$ are
the $EARL$ value for the Shewhart-MCV and RR$_{r,s}-$MCV chart. If
$\Delta_E>0$, the RR$_{r,s}-$MCV charts give better performance than the Shewhart-MCV
chart; if $\Delta_E<0$, the Shewhart-MCV chart is better.  Once again, the obtained results show that the RR$_{r,s}-$MCV charts outperforms the Shewhart-MCV chart in most cases.

\vspace{0.25cm}
In comparison with the Run Sum MCV control chart suggested by \citet{Lim2017runsum-MCV}, the Run Rules based charts also have some outstanding advantages. Indeed, to compare the performance between the Run Sum MCV chart and the Run Rules MCV chart, we present in Tables \ref{tab:tabdeltaRRvsRS}-\ref{tab:tabdeltaERRvsRS}  the values of
\begin{eqnarray}
\Delta'_A&=&100\times\frac{ARL_{RS}-ARL_{RR_{r,s}}}{ARL_{RR_{r,s}}},\\
\Delta'_E&=&100\times\frac{EARL_{RS}-EARL_{RR_{r,s}}}{EARL_{RR_{r,s}}},
\end{eqnarray}
where $ARL_{RS}$ and $EARL_{RS}$ are the $ARL$ value and the $EARL$ value of the Run Sum chart. We consider only situations where the values of $n,\gamma_0$ and $p$ in the design of the two kinds of control chart are the same. In these tables, the Run Sum chart is compared to the lower--sided RR$_{4,5}-$MCV chart for decreasing shift size ($\tau<1$) and the upper-sided RR$_{2,3}-$MCV chart for increasing shift size ($\tau>1$). 

The negative values of $\Delta'_A$ in Table \ref{tab:tabdeltaRRvsRS} show that the Run Sum chart may lead to better performance for the case a specific shift size is predetermined exactly. In contrast, the positive values of $\Delta'_E$ in Table \ref{tab:tabdeltaERRvsRS} show the better performance of the Run Rules control chart. It should be considred that in this case, the Run Sum chart was desinged based on optimizing parameters over an anticipated interval, while the Run Rules charts do not require any prediction. Even so, the global performance of the Run Rules charts is still better than that of the Run Sum chart.
In particular, with $n=5, \gamma_0=0.1$ and $p=2$, for the upward chart we have $EARL=29.4$ in the RR$^+_{2.3}-$MCV chart (Table \ref{tab:tabEARL} in this paper) while $EARL=31.37$ in the Run Sum control chart (Table 1 in \citet{Lim2017runsum-MCV}); for the downward we have $EARL=67.9$ in the RR$^-_{4.5}-$MCV chart (Table \ref{tab:tabEARL} in this paper) while $EARL=70.49$ in the Run Sum control chart (Table 3 in \citet{Lim2017runsum-MCV}).  

As mentioned above, the use of the Run Sum control chart requires to optimize the score vectors over a range of shifts that is difficult  to predetermine exactly in practice. When the predetermined value of the shift size $\tau$ is different from the
true shift size, the run-length properties of the designed control chart could be seriously affected (\citet{tran2019performance}). Meanwhile, the Run Rules charts only need the determination of a single control limit value for all shift sizes. This makes the Run Rules MCV chart more easier to implement.  

\begin{table}[thb]
  \scalebox{0.9}{
    \begin{tabular}{cccccccccccc}
     \hline
       &  \multicolumn{11}{c}{$p=2$} \\
      \hline
  & \multicolumn{3}{c}{RR$_{2,3}$-MCV chart} & &\multicolumn{3}{c}{RR$_{3,4}$-MCV chart} && \multicolumn{3}{c}{RR$_{4,5}$-MCV chart}\\
      \cline{2-4} \cline{6-8} \cline{10-12}
$\tau$&  $n=5$ & $n=10$ & $n=15$  &&  $n=5$ & $n=10$ & $n=15$ &&  $n=5$ & $n=10$ & $n=15$ \\
     \cline{2-4} \cline{6-8} \cline{10-12}
        &  \multicolumn{11}{c}{$\gamma_0=0.1$} \\
 (D) & 38 & 41 & 41 & & 53 & 54 & 53 & & 61 & 60 & 58 \\
 (I) &  5 & 12 & 14 & &  2 & 12 & 13 & & -4 &  8 &  8 \\
      \cline{2-4} \cline{6-8} \cline{10-12}
        &  \multicolumn{11}{c}{$\gamma_0=0.2$} \\
 (D) & 38 & 41 & 41 & & 53 & 54 & 53 & & 60 & 60 & 58 \\
 (I) &  5 & 12 & 14 & &  2 & 12 & 14 & & -3 &  9 &  9 \\
     \cline{2-4} \cline{6-8} \cline{10-12}
        &  \multicolumn{11}{c}{$\gamma_0=0.3$} \\
 (D) & 38 & 41 & 41 & & 52 & 54 & 53 & & 60 & 60 & 58 \\
 (I) &  6 & 13 & 15 & &  3 & 13 & 15 & & -2 & 10 & 11 \\
     \cline{2-4} \cline{6-8} \cline{10-12}
        &  \multicolumn{11}{c}{$\gamma_0=0.4$} \\
(D) & 37 & 40 & 40 & & 52 & 53 & 52 & & 59 & 59 & 57 \\
 (I) &  7 & 14 & 16 & &  5 & 15 & 17 & &  1 & 12 & 14 \\
     \cline{2-4} \cline{6-8} \cline{10-12}
        &  \multicolumn{11}{c}{$\gamma_0=0.5$} \\
 (D) & 36 & 39 & 39 & & 51 & 52 & 51 & & 59 & 58 & 57 \\
 (I) & 10 & 16 & 17 & &  9 & 17 & 19 & &  5 & 15 & 17 \\
     \cline{2-4} \cline{6-8} \cline{10-12}
       \hline
       &  \multicolumn{11}{c}{$p=3$} \\
      \hline
        & \multicolumn{3}{c}{RR$_{2,3}$-MCV chart} & &\multicolumn{3}{c}{RR$_{3,4}$-MCV chart} && \multicolumn{3}{c}{RR$_{4,5}$-MCV chart}\\
      \cline{2-4} \cline{6-8} \cline{10-12}
$\tau$&  $n=5$ & $n=10$ & $n=15$  &&  $n=5$ & $n=10$ & $n=15$ &&  $n=5$ & $n=10$ & $n=15$ \\
     \cline{2-4} \cline{6-8} \cline{10-12}
        &  \multicolumn{11}{c}{$\gamma_0=0.1$} \\
 (D) & 34 & 41 & 41 & & 49 & 55 & 53 & & 58 & 61 & 59 \\
 (I) &  1 & 11 & 14 & & -5 & 11 & 13 & & -12 &  7 &  8 \\
     \cline{2-4} \cline{6-8} \cline{10-12}
        &  \multicolumn{11}{c}{$\gamma_0=0.2$} \\
 (D) & 34 & 41 & 41 & & 49 & 54 & 53 & & 57 & 60 & 58 \\
 (I) &  1 & 12 & 14 & & -5 & 11 & 14 & & -11 &  8 &  9 \\
     \cline{2-4} \cline{6-8} \cline{10-12}
        &  \multicolumn{11}{c}{$\gamma_0=0.3$} \\
 (D) & 33 & 41 & 41 & & 48 & 54 & 53 & & 57 & 60 & 58 \\
 (I) &  2 & 12 & 15 & & -4 & 12 & 15 & & -10 &  9 & 11 \\
     \cline{2-4} \cline{6-8} \cline{10-12}
        &  \multicolumn{11}{c}{$\gamma_0=0.4$} \\
 (D) & 33 & 40 & 40 & & 48 & 53 & 52 & & 56 & 59 & 58 \\
 (I) &  3 & 13 & 16 & & -1 & 14 & 17 & & -8 & 11 & 14 \\
     \cline{2-4} \cline{6-8} \cline{10-12}
        &  \multicolumn{11}{c}{$\gamma_0=0.5$} \\
 (D) & 32 & 39 & 39 & & 47 & 52 & 52 & & 55 & 59 & 57 \\
 (I) &  6 & 15 & 17 & &  2 & 16 & 19 & & -3 & 14 & 17 \\
     \cline{2-4} \cline{6-8} \cline{10-12}
      \hline
       &  \multicolumn{11}{c}{$p=4$} \\
      \hline
        & \multicolumn{3}{c}{RR$_{2,3}$-MCV chart} & &\multicolumn{3}{c}{RR$_{3,4}$-MCV chart} && \multicolumn{3}{c}{RR$_{4,5}$-MCV chart}\\
      \cline{2-4} \cline{6-8} \cline{10-12}
$\tau$&  $n=5$ & $n=10$ & $n=15$  &&  $n=5$ & $n=10$ & $n=15$ &&  $n=5$ & $n=10$ & $n=15$ \\
     \cline{2-4} \cline{6-8} \cline{10-12}
        &  \multicolumn{11}{c}{$\gamma_0=0.1$} \\
(D) & 23 & 41 & 41 & & 37 & 55 & 54 & & 46 & 61 & 59 \\
 (I) & -8 & 11 & 14 & & -20 & 10 & 13 & & -32 &  6 &  8 \\
     \cline{2-4} \cline{6-8} \cline{10-12}
        &  \multicolumn{11}{c}{$\gamma_0=0.2$} \\
  (D) & 23 & 41 & 41 & & 37 & 54 & 53 & & 46 & 61 & 59 \\
 (I) & -7 & 11 & 14 & & -19 & 10 & 13 & & -31 &  6 &  9 \\
     \cline{2-4} \cline{6-8} \cline{10-12}
        &  \multicolumn{11}{c}{$\gamma_0=0.3$} \\
 (D) & 22 & 40 & 41 & & 36 & 54 & 53 & & 45 & 60 & 58 \\
 (I) & -7 & 11 & 15 & & -18 & 11 & 15 & & -30 &  8 & 11 \\
     \cline{2-4} \cline{6-8} \cline{10-12}
        &  \multicolumn{11}{c}{$\gamma_0=0.4$} \\
 (D) & 22 & 40 & 40 & & 35 & 53 & 52 & & 45 & 60 & 58 \\
 (I) & -5 & 12 & 16 & & -16 & 13 & 16 & & -27 & 10 & 14 \\
     \cline{2-4} \cline{6-8} \cline{10-12}
        &  \multicolumn{11}{c}{$\gamma_0=0.5$} \\
 (D) & 21 & 39 & 39 & & 35 & 52 & 52 & & 44 & 59 & 57 \\
 (I) & -3 & 14 & 17 & & -12 & 15 & 19 & & -22 & 13 & 17 \\
\hline
    \end{tabular}}
          \caption{The $\Delta_E$ index values for different values of $p, n$ and $\gamma_0$.}
  \label{tab:tabDeltaEARL}
  \end{table}

   \begin{table}[thb]
  \hspace*{20mm}
  \scalebox{1}{
    \begin{tabular}{cccccc}
     \hline
     &  \multicolumn{2}{c}{$p=2$} && \multicolumn{2}{c}{$p=3$}\\
      \cline{2-3} \cline{5-6} 
$\tau$&  $n=5$ & $n=10$ && $n=5$ & $n=10$\\
\cline{2-6}
       &  \multicolumn{5}{c}{$\gamma_0=0.1$} \\
0.50	&	-10	&	-36	&&	-22	&	-30	\\
0.75	&	-32	&	-15	&&	-37	&	-17	\\
0.90	&	-21	&	-20	&&	-21	&	-20	\\
1.10	&	-19	&	-27	&&	-17	&	-26	\\
1.25	&	-27	&	-29	&&	-25	&	-29	\\
1.50	&	-26	&	-22	&&	-27	&	-22	\\
    \cline{2-6} 
        &  \multicolumn{5}{c}{$\gamma_0=0.3$} \\
0.50	&	-10	&	-33	&&	-24	&	-28	\\
0.75	&	-32	&	-16	&&	-37	&	-18	\\
0.90	&	-21	&	-20	&&	-20	&	-20	\\
1.10	&	-19	&	-26	&&	-16	&	-24	\\
1.25	&	-27	&	-29	&&	-25	&	-29	\\
1.50	&	-26	&	-22	&&	-27	&	-23	\\
     \cline{2-6} 
        &  \multicolumn{5}{c}{$\gamma_0=0.5$} \\
0.50	&	-11	&	-29	&&	-26	&	-23	\\
0.75	&	-31	&	-18	&&	-36	&	-20	\\
0.90	&	-19	&	-19	&&	-19	&	-19	\\
1.10	&	-17	&	-23	&&	-14	&	-23	\\
1.25	&	-25	&	-30	&&	-23	&	-30	\\
1.50	&	-25	&	-24	&&	-25	&	-24	\\

\hline
    \end{tabular}}
      \caption{The values of  $\Delta'_A$ index for $p\in\{2,3\}$, $n\in\{5,10\}$ and  $\gamma_0\in\{0.1,0.3,0.5\}$.}
  \label{tab:tabdeltaRRvsRS}
  \end{table}
  
   \begin{table}[thb]
  \hspace*{20mm}
  \scalebox{1}{
   \begin{tabular}{cccccc}
     \hline
     &  \multicolumn{2}{c}{$p=2$} && \multicolumn{2}{c}{$p=3$}\\
      \cline{2-3} \cline{5-6} 
$\tau$&  $n=5$ & $n=10$ && $n=5$ & $n=10$\\
\cline{2-6}
        &  \multicolumn{5}{c}{$\gamma_0=0.1$} \\
(D)&	4	&	30		&&	-5	&	27	\\
(I)&	6	&	18		&&	3	&	16	\\
    \cline{2-6} 

        &  \multicolumn{5}{c}{$\gamma_0=0.3$} \\
(D)&	3	&	29		&&	-6	&	25	\\
(I)&	5	&	16		&&	2	&	14	\\
     \cline{2-6} 
        &  \multicolumn{5}{c}{$\gamma_0=0.5$} \\
(D)&	2	&	25		&&	-6	&	22	\\
(I)&	3	&	11		&&	0	&	9	\\

\hline
    \end{tabular}}
      \caption{The $\Delta'_E$ index values   for $p\in\{2,3\}$, $n\in\{5,10\}$ and  $\gamma_0\in\{0.1,0.3,0.5\}$.}
  \label{tab:tabdeltaERRvsRS}
  \end{table}
  
\section{Illustrative example}
\label{sec:illustrative}
An illustrative example of RR$_{r,s}-$MCV control chart is given in this Section. Let us consider a sintering process in an Italian company that manufactures sintered mechanical parts, which is introduced in \citet{Lim2017runsum-MCV}. The data are recorded from a spring manufacturing process, for which the quality characteristics are the spring inner diameter ($X_1$) and the spring elasticity ($X_2$). From Phase I, we have the estimated value of $\gamma_0=0.089115$, while (according to \citet{Lim2017runsum-MCV}) the assumption that the MCV during Phase I is constant holds. The data collected during the Phase II process  with sample size $n=5$ are shown in Table \ref{tab:data}. Further details on the process can be found in \citet{Lim2017runsum-MCV}. From the obtained results in Section \ref{sec:implementationnonME}, the control limits for the different control charts are as follows.
\begin{itemize}
\item For the upper-sided Shewhart$-$MCV chart, $UCL_{\mathrm{Shewhart}}=0.1691.$
\item For the upper-sided RR$^+_{2,3}-\mathrm{MCV}$ chart, $UCL_{\mathrm{RR}_{2,3}}= 0.1296.$
\item For the upper-sided RR$^+_{3,4}-\mathrm{MCV}$ chart, $UCL_{\mathrm{RR}_{3,4}} = 0.1106.$
\item For the upper-sided RR$^+_{4,5}-\mathrm{MCV}$ chart, $UCL_{\mathrm{RR}_{4,5}} = 0.0986.$
\end{itemize}

\begin{table}[thb]
\renewcommand{\baselinestretch}{1}\small\normalsize
\begin{center}
\scalebox{0.9}{
\begin{tabular}{ccccccc}
    \hline
   Sample number $t$ &$\bar{X}_{1,t}$& $\bar{X}_{2t}$& $S^2_{1t}$& $S^2_{2t}$& $S_{12t}$&  $\hat{\gamma}_i$ \\
    \hline  
1& 7.781& 1.592& 1.164& 0.734 &0.35645& 0.113710\\
2& 7.385& 1.804& 1.006& 1.667& 0.96049 &0.104890\\
3& 7.988& 2.260& 0.762& 0.359& 0.17373& 0.108870\\
4& 8.189& 2.100& 1.885& 0.470& 0.13026& 0.156790\\ 
5& 7.436& 2.061& 1.404& 0.519 &0.08280& 0.139290\\ 
6& 6.746& 2.289& 0.846& 0.811& 0.43835& 0.133240\\ 
7& 7.356& 1.917& 0.197& 2.587&0.01597& 0.059996\\ 
8& 8.492& 1.845& 1.460& 1.746&1.42051& 0.055093\\ 
9& 7.272& 1.580& 1.353& 0.345&0.27988& 0.117710\\ 
10&7.585& 1.568& 1.098&0.788&0.41252&0.109610\\
11& 7.734& 1.709& 0.952& 0.228 &0.11462& 0.102440\\
12& 8.160& 1.498& 1.598& 1.178& 1.00757& 0.122950\\ 
13& 7.102& 2.661& 1.508& 0.945 &0.73607& 0.101260\\
14& 8.392& 1.883& 0.536& 0.706& 0.23234& 0.085637\\ 
15& 7.592& 2.531& 0.256& 0.563&0.24827& 0.043489\\ 
16& 8.141& 2.093& 0.394& 0.603& 0.25584& 0.072202\\ 
17& 7.883& 2.490& 1.321& 1.179& 0.65037& 0.142430 \\
18 &7.886& 2.877& 0.883& 1.431 &0.22524& 0.106680\\
19& 7.830& 1.008& 0.878& 0.558 &0.14223& 0.112090\\ 
20& 8.196& 1.482& 0.791& 0.220&0.13724& 0.088460\\
    \hline
  \end{tabular}}
\end{center}
  \caption{Illustrative example of Phase II dataset.}
  \label{tab:data}
\renewcommand{\baselinestretch}{1.4}\small\normalsize
\end{table}

\begin{figure}[thb]
  \begin{center}
    \includegraphics[width=100mm]{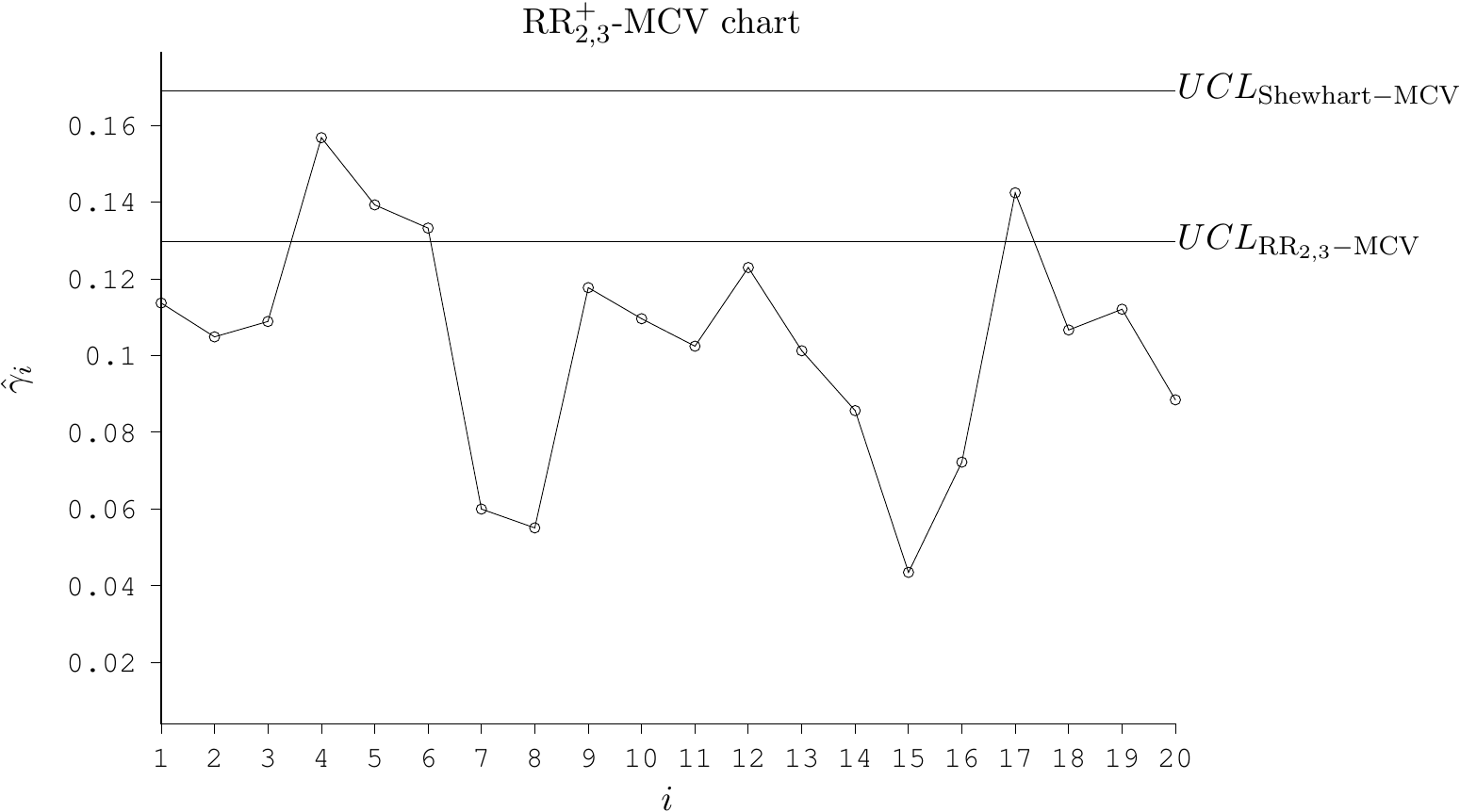} \\[5mm]
  \end{center}
 \caption{ RR$^+_{2,3}-$MCV control chart corresponding to Phase II data set in Table~\ref{tab:data}}
\label{fig:mcvchart1}
\end{figure}

\begin{figure}[thb]
  \begin{center}
    \includegraphics[width=100mm]{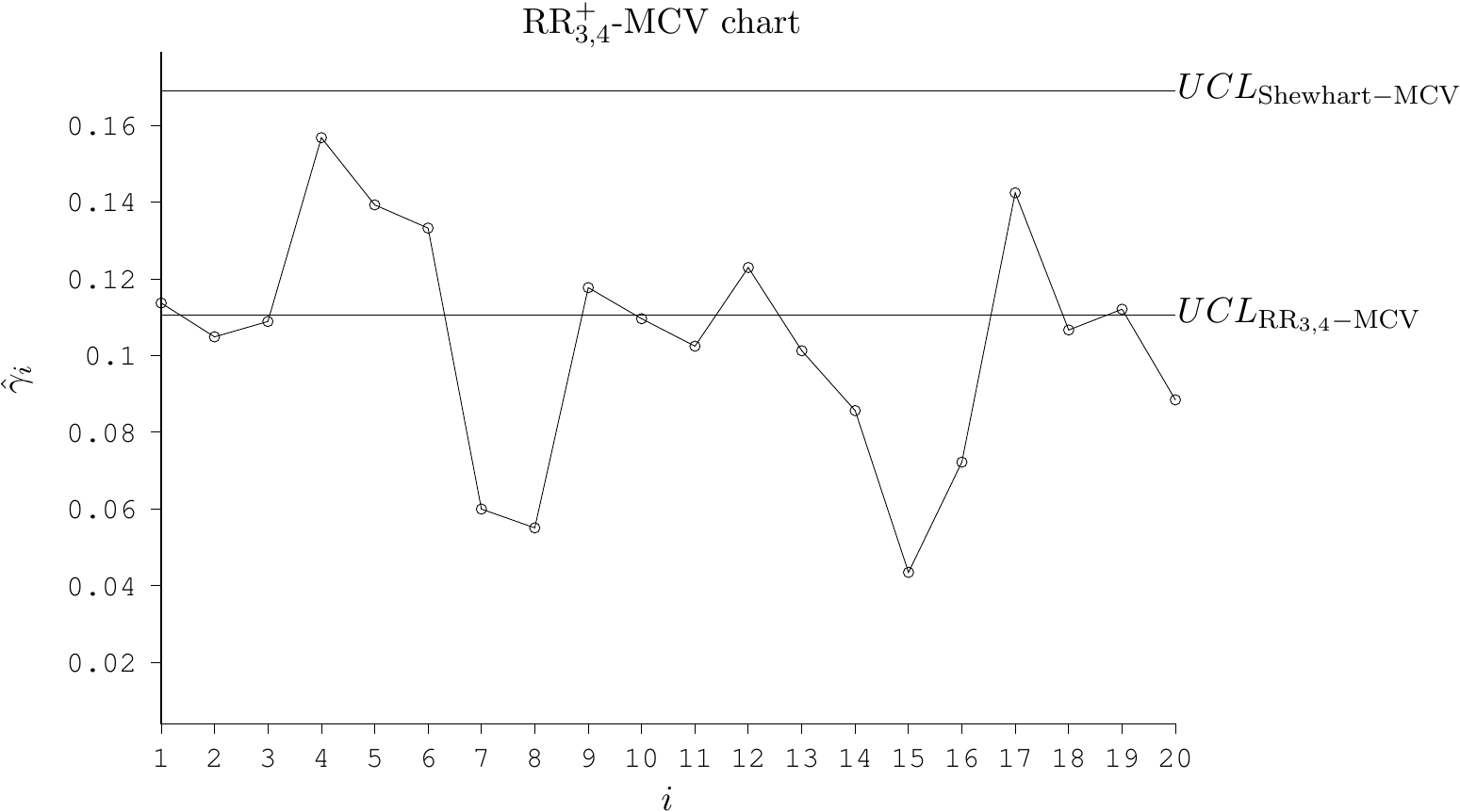} \\[5mm]
  \end{center}
  \caption{ RR$^+_{3,4}-$MCV control chart corresponding to Phase II data set in Table~\ref{tab:data}}
\label{fig:mcvchart2}
\end{figure}

\begin{figure}[thb]
  \begin{center}
    \includegraphics[width=100mm]{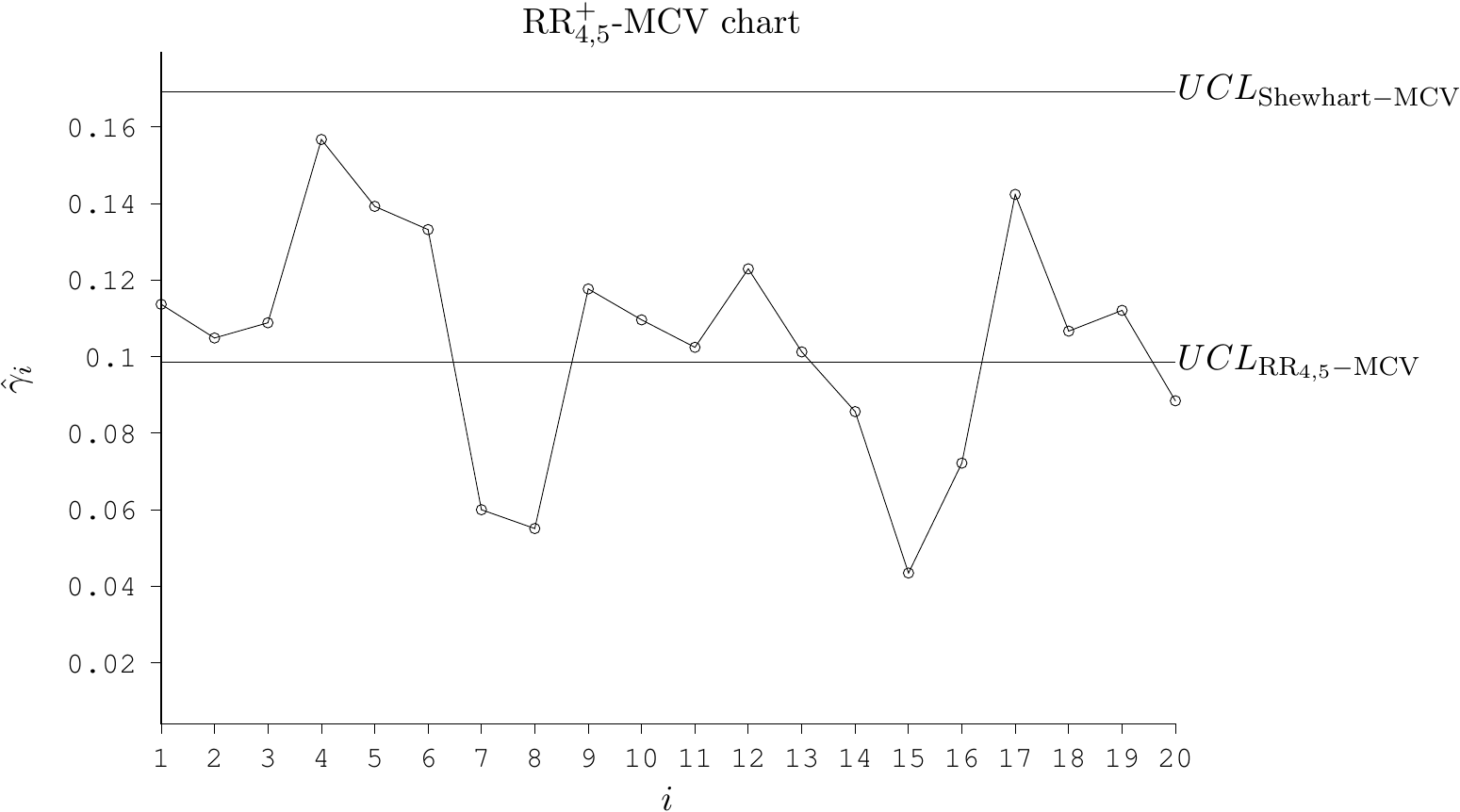} \\[5mm]
  \end{center}
  \caption{ RR$^+_{4,5}-$MCV control chart corresponding to Phase II data set in Table \ref{tab:data}}
\label{fig:mcvchart3}
\end{figure}

The corresponding values of $\hat{\gamma}_i$ are presented in the rightmost column of Table \ref{tab:data} and plotted
in Figures~\ref{fig:mcvchart1}-\ref{fig:mcvchart3}, respectively. Each figure consists of the $UCL$ for the upper-sided Shewhart - MCV chart  along with the $UCL$ of the $RR^{+}_{r,s}-$MCV control chart, for $(r,s)\in\{(2,3),(3,4),(4,5)\}$. It is not difficult to see that the  RR$^+_{2,3}-$MCV
control chart signals the occurrence of the out-of-control condition by two out of three successive plotting points \#4 and \#5 above
the control limit $UCL_{\mathrm{RR}_{2,3}}$, see Figure~\ref{fig:mcvchart1}.
The RR$_{3,4}-$MCV control chart signals the occurrence of the out-of-control condition by three out of four successive plotting points \#4, \#5 and \#6 above
the control limit $UCL_{\mathrm{RR}_{3,4}}$, see Figure~\ref{fig:mcvchart2}.
The RR$_{4,5}-$MCV control chart signals the occurrence of the out-of-control condition by four out of five
successive plotting points \#1, \#2, \#3 and \#4 above
the control limit $UCL_{\mathrm{RR}_{4,5}}$, see Figure~\ref{fig:mcvchart3}.  The
 figures~\ref{fig:mcvchart1}-\ref{fig:mcvchart3} also show that all the points are plotted below the $UCL_{Shewhart}$. That is to say, in this situation, the upper Shewhart - MCV control chart (designed in \citet{Yeong_MCV_2016}) fails to detect the out-control condition detected by the RR$_{r,s}-$MCV control charts. In addition, it is worth noting that similar conclusions have been reached by \citet{Lim2017runsum-MCV} with the Run Sum MCV control chart. Specifically, their Run Sum chart gives an out-of-control signal at sample 4. However, the design of this run sum control chart requires the \textit{a priori} selection for a shift of interest so as to determine the optimal scores for the chart. This means that the run sum chart is designed so as to be the optimal one in the selection of a specific shift. Their choice, for illustrative purposes, was a 25\% increase in the IC value of the MCV. Clearly, this is not required here where only the value of the $UCL$ is needed for the application of the Run Rules chart. Before closing this section we mention that the lower-sided $RR^{-}_{r,s}-$ MCV control charts can be constructed in a similar manner. The respective limits are 0.010029, 0.02403, 0.03464, 0.04275 . As in the case of the lower-sided run sum chart of \citet{Lim2017runsum-MCV} no out-of-control signal is given. However, due to space constraints, we do not provide the respective figures.
\section{Concluding remarks}
\label{sec:conclusions}
In this paper, we have investigated the one-sided control charts with Run Rules for monitoring the coefficient of variation of multivariate data. Two one-sided charts were considered to detect separately both  increases and  decreases in the multivariate CV. The performance of proposed charts is evaluated  through $ARL$ for deterministic shift size and $EARL$ for unknown shift size. The numerical results showed that the Run Rules control charts enhance the performance of Shewhart control chart significantly. For purpose of optimizing the performance of Run Rules charts, it is recommended to use the RR$^-_{4,5}-$MCV for detecting decreasing process shifts and  RR$^+_{2,3}-$MCV for detecting  increasing process shifts. 
Moreover, under certain conditions, this careful choice of the Run Rules charts also lead to an improved efficiency compared to the Run Sum control chart for MCV.\\

%


\begin{thebibliography}{35}
\providecommand{\natexlab}[1]{#1}
\providecommand{\url}[1]{\texttt{#1}}
\expandafter\ifx\csname urlstyle\endcsname\relax
  \providecommand{\doi}[1]{doi: #1}\else
  \providecommand{\doi}{doi: \begingroup \urlstyle{rm}\Url}\fi

\bibitem[Castagliola et~al.(2011)Castagliola, Amdouni, Taleb, and
  Celano]{Castagliola_EWMA_CV_2011}
P.~Castagliola, A.~Amdouni, H.~Taleb, and G.~Celano.
\newblock Monitoring the coefficient of variation using ewma charts.
\newblock \emph{Journal of Quality Technology}, 43\penalty0 (3):\penalty0
  249--265, 2011.

\bibitem[Kang et~al.(2007)Kang, Lee, Seong, and Hawkins]{Kang_CV_2007}
C.W. Kang, M.S. Lee, Y.J. Seong, and D.M. Hawkins.
\newblock A control chart for the coefficient of variation.
\newblock \emph{Journal of Quality Technology}, 39\penalty0 (2):\penalty0
  151--158, 2007.

\bibitem[Calzada and Scariano(2013)]{Calzada2013}
M.E. Calzada and S.~M. Scariano.
\newblock A synthetic control chart for the coefficient of variation.
\newblock \emph{Journal of Statistical Computation and Simulation}, 83\penalty0
  (5):\penalty0 853--867, 2013.

\bibitem[Tran and Tran(2016)]{Tranhanh2016_CUSUM_CV}
P.H. Tran and K.~P. Tran.
\newblock The efficiency of {CUSUM} schemes for monitoring the coefficient of
  variation.
\newblock \emph{Applied Stochastic Models in Business and Industry},
  32\penalty0 (6):\penalty0 870--881, 2016.

\bibitem[Castagliola et~al.(2013{\natexlab{a}})Castagliola, Achoure, Taleb,
  Celano, and Psarakis]{Castagliola_Run-Rules_CV_2013}
P.~Castagliola, A.~Achoure, H.~Taleb, G.~Celano, and S.~Psarakis.
\newblock Monitoring the coefficient of variation using control charts with run
  rules.
\newblock \emph{Quality Technology \& Quantitative Management}, 10:\penalty0
  75--94, 2013{\natexlab{a}}.

\bibitem[Castagliola et~al.(2013{\natexlab{b}})Castagliola, Achouri, Taleb,
  Celano, and Psarakis]{CastagliolaVSI_CV_2013}
P.~Castagliola, A.~Achouri, H.~Taleb, G.~Celano, and S.~Psarakis.
\newblock Monitoring the coefficient of variation using a variable sampling
  interval control chart.
\newblock \emph{Quality and Reliability Engineering International}, 29\penalty0
  (8):\penalty0 1135--1149, 2013{\natexlab{b}}.

\bibitem[You et~al.(2016)You, Khoo, Castagliola, and Haq]{You2015_CV_sensitive}
H.W. You, B.C.M.~Michael Khoo, P.~Castagliola, and A.~Haq.
\newblock Monitoring the coefficient of variation using the side sensitive
  group runs chart.
\newblock \emph{Quality and Reliability Engineering International}, 32\penalty0
  (5):\penalty0 1913--1927, 2016.

\bibitem[Teoh et~al.(2017)Teoh, Khoo, Castagliola, Yeong, and
  Teh]{Teoh2017Runsum_CV}
W.L. Teoh, M.B. Khoo, P.~Castagliola, W.C. Yeong, and S.Y. Teh.
\newblock Run-sum control charts for monitoring the coefficient of variation.
\newblock \emph{European Journal of Operational Research}, 257\penalty0
  (1):\penalty0 144--158, 2017.

\bibitem[Zhang et~al.(2014)Zhang, Li, Chen, and Wang]{Zhang2014}
J.~Zhang, Z.~Li, B.~Chen, and Z.~Wang.
\newblock A new exponentially weighted moving average control chart for
  monitoring the coefficient of variation.
\newblock \emph{Computers \& Industrial Engineering}, 78:\penalty0 205--212,
  2014.

\bibitem[Yeong et~al.(2017)Yeong, Khoo, Teoh, and
  Rahim]{Yeong_VSI_EWMA_CV_2017}
W.C. Yeong, M.B.C. Khoo, L.K. Teoh, and M.A. Rahim.
\newblock Monitoring the coefficient of variation using a variable sampling
  interval ewma chart.
\newblock \emph{Journal of Quality Technology}, 49\penalty0 (3):\penalty0
  380--401, 2017.

\bibitem[Zhang et~al.(2018)Zhang, Li, and Wang]{zhang2018control}
Jiujun Zhang, Zhonghua Li, and Zhaojun Wang.
\newblock Control chart for monitoring the coefficient of variation with an
  exponentially weighted moving average procedure.
\newblock \emph{Quality and Reliability Engineering International}, 34\penalty0
  (2):\penalty0 188--202, 2018.

\bibitem[Shu and Jiang(2008)]{shu2008new}
Lianjie Shu and Wei Jiang.
\newblock A new ewma chart for monitoring process dispersion.
\newblock \emph{Journal of Quality Technology}, 40\penalty0 (3):\penalty0
  319--331, 2008.

\bibitem[Albert and Zhang(2010)]{albert2010novel}
A.~Albert and L.~Zhang.
\newblock A novel definition of the multivariate coefficient of variation.
\newblock \emph{Biometrical Journal}, 52\penalty0 (5):\penalty0 667--675, 2010.

\bibitem[Yeong et~al.(2016)Yeong, Khoo, L.Teoh, and
  Castagliola]{Yeong_MCV_2016}
W.C. Yeong, M.~B.~C. Khoo, W.~L.Teoh, and P.~Castagliola.
\newblock A control chart for the multivariate coefficient of variation.
\newblock \emph{Quality and Reliability Engineering International}, 32\penalty0
  (3):\penalty0 1213--1225, 2016.

\bibitem[Lim et~al.(2017)Lim, Khoo, Teoh, and Haq]{Lim2017runsum-MCV}
A.~Lim, M.B.C Khoo, W.L. Teoh, and A.~Haq.
\newblock Run sum chart for monitoring multivariate coefficient of variation.
\newblock \emph{Computers \& Industrial Engineering}, 109:\penalty0 84--95,
  2017.

\bibitem[Champ and Woodall(1987)]{champ1987exact}
C.W. Champ and W.H. Woodall.
\newblock Exact results for shewhart control charts with supplementary runs
  rules.
\newblock \emph{Technometrics}, 29\penalty0 (4):\penalty0 393--399, 1987.

\bibitem[Klein(2000)]{Klein2000}
M.~Klein.
\newblock Two {A}lternatives to the {S}hewhart $\bar{X}$ {C}ontrol {C}hart.
\newblock \emph{Journal of Quality Technology}, 32:\penalty0 427--431, 2000.

\bibitem[Khoo(2004)]{Khoo2004}
M.B.C. Khoo.
\newblock Design of {R}uns {R}ules {S}chemes.
\newblock \emph{Quality Engineering}, 16\penalty0 (2):\penalty0 27--43, 2004.

\bibitem[Antzoulakos and Rakitzis(2008)]{Antzoulakos2008}
D.L. Antzoulakos and A.C. Rakitzis.
\newblock The {M}odified $r$ out of $m$ {C}ontrol {C}hart.
\newblock \emph{Communications in Statistics -- Simulation and Computation},
  37\penalty0 (2):\penalty0 396--408, 2008.

\bibitem[Koutras et~al.(2007)Koutras, Bersimis, and Maravelakis]{Koutras_2007}
M.V. Koutras, S.~Bersimis, and P.~Maravelakis.
\newblock Statistical process control using shewhart control charts with
  supplementary runs rules.
\newblock \emph{Methodology and Computing in Applied Probability}, 9:\penalty0
  207--224, 2007.

\bibitem[Acosta and Pignatiello(2009)]{Acosta_2009_kofkRR}
C.~Acosta and J.~Pignatiello.
\newblock Arl-design of s charts with k-of-k runs rules.
\newblock \emph{Communications in Statistics - Simulation and Computation},
  38:\penalty0 1625--1639, 2009.

\bibitem[Amdouni et~al.(2016)Amdouni, Castagliola, Taleb, and
  Celano]{Amdouni_2016_one-sided-RR}
A.~Amdouni, P.~Castagliola, H.~Taleb, and G.~Celano.
\newblock One-sided run rules control charts for monitoring the coefficient of
  variation in short production runs.
\newblock \emph{European Journal of Industrial Engineering}, 10\penalty0
  (5):\penalty0 639--663, 2016.

\bibitem[Faraz et~al.(2014)Faraz, Celano, Saniga, Heuchenne, and
  Fichera]{Faraz_2014_T2RR}
A.~Faraz, G.~Celano, E.~Saniga, C.~Heuchenne, and S.~Fichera.
\newblock The variable parameters t$^{2} $ chart with run rules.
\newblock \emph{Statistical Papers}, 55\penalty0 (4):\penalty0 933--950, 2014.

\bibitem[Tran et~al.(2016)Tran, Castagliola, and Celano]{Tran2016_Runrules_RZ}
K.P. Tran, P.~Castagliola, and G.~Celano.
\newblock Monitoring the {R}atio of {T}wo {N}ormal {V}ariables {U}sing {R}un
  {R}ules {T}ype {C}ontrol {C}harts.
\newblock \emph{International Journal of Production Research}, 54\penalty0
  (6):\penalty0 1670--1688, 2016.

\bibitem[Chew et~al.(2019)Chew, Khaw, and Yeong]{chew2019efficiency}
X.~Y. Chew, K.~W. Khaw, and W.~C. Yeong.
\newblock The efficiency of run rules schemes for the multivariate coefficient
  of variation: a markov chain approach.
\newblock \emph{Journal of Applied Statistics}, pages 1--21, 2019.

\bibitem[Tran(2016)]{Tran2016_4-out-of-5-Runrules_RZ}
K.P. Tran.
\newblock The efficiency of the 4-out-of-5 {R}uns {R}ules scheme for monitoring
  the {R}atio of {P}opulation {M}eans of a {B}ivariate {N}ormal distribution.
\newblock \emph{International Journal of Reliability, Quality and Safety
  Engineering}, 2016.
\newblock In press, DOI: 10.1142/S0218539316500200.

\bibitem[Reyment(1960)]{Reyment1960}
R.A. Reyment.
\newblock Studies on nigerian upper cretaceous and lower tertiary ostracoda.
  part 1. senonian and maestrichtian ostracoda 1960.
\newblock \emph{Stockholm Contributions in Geology}, 7:\penalty0 1?238, 1960.

\bibitem[Valen(1974)]{van1974multivariate}
L.~Van Valen.
\newblock Multivariate structural statistics in natural history.
\newblock \emph{Journal of Theoretical Biology}, 45\penalty0 (1):\penalty0
  235--247, 1974.

\bibitem[Nikulin and Voinov(2011)]{nikulin2011unbiased}
M.~Nikulin and V.~Voinov.
\newblock Unbiased estimators and their applications.
\newblock In \emph{International Encyclopedia of Statistical Science}, pages
  1619--1621. Springer, 2011.

\bibitem[Brook and Evans(1972)]{Brook1972}
D.~Brook and D.A. Evans.
\newblock An approach to the probability distribution of {CUSUM} run length.
\newblock \emph{Biometrika}, 59\penalty0 (3):\penalty0 539--549, 1972.

\bibitem[Fu et~al.(2003)Fu, Spiring, and Xie]{Fu_2003_Markov_approach}
J.~Fu, F.~Spiring, and H.~Xie.
\newblock On the average run lengths of quality control schemes using a markov
  chain approach.
\newblock \emph{Statistics \& Probability Letters}, 56:\penalty0 369--380.,
  2003.

\bibitem[Li et~al.(2014)Li, Zou, Gong, and Wang]{Li2014}
Z.~Li, C.~Zou, Z.~Gong, and Z.~Wang.
\newblock The computation of average run length and average time to signal: an
  overview.
\newblock \emph{Journal of Statistical Computation and Simulation}, 84\penalty0
  (8):\penalty0 1779--1802, 2014.

\bibitem[Chen and Chen(2007)]{Chen2007b}
A.~Chen and Y.~K. Chen.
\newblock Design of {EWMA} and {CUSUM} control charts subject to random shift
  sizes and quality impacts.
\newblock \emph{IIE Transactions}, 39\penalty0 (12):\penalty0 1127--1141, 2007.

\bibitem[Celano et~al.(2014)Celano, Castagliola, Faraz, and
  Fichera]{Celano2014a}
G.~Celano, P.~Castagliola, A.~Faraz, and S.~Fichera.
\newblock Statistical {P}erformance of a {C}ontrol {C}hart for {I}ndividual
  {O}bservations {M}onitoring the {R}atio of two {N}ormal {V}ariables.
\newblock \emph{Quality and Reliability Engineering International}, 30\penalty0
  (8):\penalty0 1361--1377, 2014.

\bibitem[Tran et~al.(2019)Tran, Nguyen, Tran, and
  Heuchenne]{tran2019performance}
K.~P. Tran, H.~D. Nguyen, P.~H. Tran, and C.~Heuchenne.
\newblock On the performance of cusum control charts for monitoring the
  coefficient of variation with measurement errors.
\newblock \emph{The International Journal of Advanced Manufacturing
  Technology}, pages 1--15, 2019.

\end{thebibliography}

\end{document}